\documentstyle[12pt]{article}
\newskip\zatskip \zatskip=0pt plus0pt minus0pt
\def\matth{\mathsurround=0pt}

\def\gsim{\mathrel{\mathpalette\atversim>}}
\def\atversim#1#2{\lower0.7ex\vbox{\baselineskip\zatskip\lineskip\zatskip
  \lineskiplimit 0pt\ialign{$\matth#1\hfil##\hfil$\crcr#2\crcr\sim\crcr}}}

\input epsf

\begin{document}

\begin{titlepage}



\let\picnaturalsize=N
\def\picsize{1.0in}
\def\picfilename{scipp_tree.eps}


\ifx\nopictures Y\else
{\hbox to\hsize{\hbox{\ifx\picnaturalsize N\epsfxsize \picsize\fi
{\epsfbox{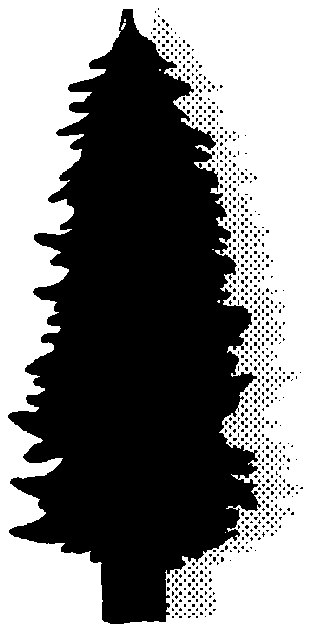}}}\hspace*{\fill}

\parbox[b]{3.4cm}{SCIPP 95/14 \\ April 1995\\ hep-ph/9504434\\
\vspace*{2.0cm}}

}}\fi


\vspace*{-0.5cm}

\begin{center}
\large\bf
Top Quark Production and Decay at Next-to-leading Order
in $e^+e^-$ Annihilation
\end{center}

\vspace*{0.3cm}

\begin{center}
Carl R. Schmidt \footnote{Supported in part by the U.S.
Department of Energy.} \\
Santa Cruz Institute for Particle Physics\\
University of California, Santa Cruz, CA 95064, USA
\end{center}

\vspace*{0.3cm}

\begin{center}
\bf Abstract
\end{center}

\noindent

We study the effects of QCD corrections to the process
$e^+e^-\rightarrow t\bar t+X\rightarrow b\ell^+\nu
\bar b\ell^-\bar \nu+X$ above threshold.  We show how to treat
consistently to ${\cal O}(\alpha_s)$ the gluon radiation in both the
production and the decay of the top quarks, while maintaining all
angular correlations in the event.  At this order there is an
ambiguity in the event reconstruction whenever a real gluon occurs in
the final state.
We study the effects of this ambiguity on the top mass and helicity
angle distributions.  For a top mass of 175 GeV and collider energy of
400 GeV the gluon radiation is emitted predominantly in the decay of
the top quarks.

\end{titlepage}

\baselineskip=0.8cm

\section{Introduction}

Recently, the CDF\cite{CDF} and the D0\cite{D0} collaborations at
Fermilab announced the observation of the top quark in $p\bar p$
collisions at the Tevatron.  Both groups saw a
statistically significant excess of dilepton and lepton+jets events
with the proper kinematic properties and bottom quark tags needed to
indicate $t\bar t$ production.  Furthermore, they were able to extract
mass values for the top quark by fitting to events consisting of a
single lepton plus four jets.  The D0 group found a mass of
$199^{+19}_{-21}\pm22$ GeV, while CDF obtained a mass of $176\pm8\pm10$
GeV.  Both of these mass measurements are in excellent agreement with
the value of $175\pm11^{+17}_{-19}$ GeV obtained indirectly
from a global fit\cite{ErLang} to the electroweak data from LEP and
SLAC.  The direct observation of the top quark at the Tevatron heralds
the start of a new era in the study of flavor physics.

The top quark is certainly unique among the six known quarks.  It is
by far the heaviest; more than 30
times as massive as the bottom quark and even more massive than the
$W$ and $Z$ bosons.  Correspondingly, the top quark also has the
largest coupling to the symmetry breaking sector of all the known
particles.  This large coupling to the Higgs sector may give rise to
deviations from its expected behavior, thereby offering clues to
symmetry breaking, fermion mass generation, quark family replication,
and other deficiencies of the Standard Model.
For example, in top-color and extended technicolor (ETC)
models the top quark may have non-standard couplings to the weak
vector bosons\cite{sekhar} or
there may be resonant enhancement of $t\bar t$ production\cite{parke}.
It is of utmost importance to examine the top quark properties
as precisely as possible.

A more basic consequence of the large top quark mass is its short
lifetime.  For large mass the lifetime of the top quark scales as
$[1.7\ {\rm GeV}\cdot (m_t/175 \ {\rm
GeV})^3]^{-1}$, and so the top quark decays very rapidly
to a bottom quark and a $W$.  Thus, unlike the lighter quarks which
form hadronic bound states before decaying, the top quark behaves
more like a heavy lepton, decaying as an unbound fermion.
In fact, it decays long before depolarization, so that its spin
information can be easily reconstructed from the momenta of its decay
products.  This fact will be extremely useful for extracting
information about the top quark parameters.

An ideal place to study the top quark is in $e^+e^-$
collisions\cite{NLC,jezabek},  where
the colorless initial state provides a clean event environment, and
there is the possibility of initial-state polarization.  By
varying the beam energy it is possible to scan the threshold region or
to study the top above threshold.  There have been many studies of top
production near threshold, where the resonance behavior can be
calculated in perturbative QCD and the top mass can be obtained to high
accuracy\cite{threshold}.
In this paper we will instead concentrate on the continuum
$ t\bar t$ production.  At tree level the event
is characterized by six final-state particles arising from
the process $e^+e^-\rightarrow t\bar t
\rightarrow bW^+\bar bW^-\rightarrow b\ell^+\nu\bar b\ell^-\bar\nu$.
These six particles contain a wealth of information in their relative
momenta, angles, and polarizations.
By reconstructing the helicity angles of the top quarks and
the $W$'s, it is straightforward to extract the top quark parameters.

Although the top quark is produced and decays essentially as an
unbound fermion, it still feels the strong interactions and will
radiate gluons---both in its production phase and its decay phase.
Thus, it is useful to see how the tree-level picture and
experimental analysis will be affected by QCD corrections.  The ${\cal
O}(\alpha_s)$ corrections to the production have been studied in
several papers, including analyses of the effects on
production angle distributions\cite{JLZ} and polarizations\cite{KPT}.
Similarly, studies of
the ${\cal O}(\alpha_s)$ corrections on the top decay have been done,
with analyses of energy distributions, and angular distributions from
polarized tops\cite{JezKuhn}.  However, the top production and decays
do not occur in isolation from each other.  For events with an extra
gluon jet it is not {\it a priori} obvious whether to assign the extra
jet to the production, the $t$ decay, or the $\bar t$ decay.
At the very least, the extra jets will add one more degree of
complexity to the event reconstruction process.  Therefore, it is
necessary to assess the impact of these radiative corrections on the
full event\cite{mrcp}.

To this end we have constructed a next-to-leading order (NLO) Monte Carlo
which treats consistently to ${\cal O}(\alpha_s)$ the radiative
corrections to both production and decay of the top quarks.
To set the stage for this NLO analysis we begin by reviewing the salient
features of the $e^+e^-\rightarrow t\bar t$ event at tree level using
helicity decomposition in section 2.
Then in section 3 we analyze the cross section at
next-to-leading order and give the details of the Monte Carlo,
describing the approximations used and the methods for subtracting the
infrared (IR) divergences in production and decay.  We also include
two appendices with the helicity amplitudes for top production and
decay with real gluons.  In section 4
we use the Monte Carlo to study the effects of gluon radiation on the top
quark mass measurement and to re-examine the helicity angle
distributions at next-to-leading order.  In this section we
assume that the only ambiguities are in the placement of the extra
gluon jet, that the $W$'s and bottom quarks are correctly identified,
and we investigate how the distributions vary with the algorithm
used for assigning the gluon jet.  Then in section 5 we make another
pass through the mass distributions with more
realistic experimental assumptions for the event.
The purpose of this section is to identify which physical inputs have the
largest effect on the continuum measurement of the top mass.
In section 6 we
offer our conclusions.

\section{Review of the tree-level analysis}

Even at tree level the full $e^+e^-\rightarrow t\bar t$ event is
quite complex.  The six-particle final state can be characterized in
many possible ways by the relative momenta and angles in the event.
It is an important conceptual problem to clarify which pieces of
information are most important, and how all of the various
kinematic measurements available cooperate to illuminate the basic
physics.  The solution to this problem is suggested by the fact that
the event is actually a series of on-shell two-body decays:
$\gamma^*,Z^*\rightarrow t\bar t$, $t\rightarrow b W^+$, and $W^+\rightarrow
\ell^+\nu$.  Thus, by considering intermediate states of definite
helicities, the event is highly constrained simply by conservation of
angular momentum.  The different helicity states are revealed by the
angular distributions of their decay products, while
the relative amplitudes for the different helicity
combinations are easily related to the couplings at the top quark
production and decay vertices.  In this section we describe this
tree-level helicity analysis.  Although this has been discussed before
in the literature, most notably by Kane, Ladinsky, and Yuan\cite{KLY},
we will review it here for pedagogical purposes and to set the notation
for the discussion of QCD corrections.

The dominant effects of new physics on the process $e^+e^-\rightarrow
t\bar t\rightarrow bW^+\bar bW^-$ can be described in terms of form
factors included at the production and decay vertices.
The $t \rightarrow bW^+$ decay vertex can be written
\begin{equation}
i{\cal M}^{W\mu} = i{g\over\sqrt{2}}
  \Bigl\lbrace
 \gamma^\mu  [F^W_{1L}P_L + F^W_{1R}P_R]
 +{i\sigma^{\mu\nu} q_\nu\over2m_t}[F^W_{2L}P_R +
 F^W_{2R}P_L]
  \Bigr\rbrace\ ,\label{decayffs}
\end{equation}
where $P_{R,L}=(1\pm\gamma_5)/2$, and we have neglected a third pair of
form factors which do not contribute to decays to on-shell $W$'s or
massless fermions.  We have chosen the subscripts $L,R$ of the form
factors so that they indicate the helicity of the outgoing bottom
quark in the limit $m_b=0$, which we will use in all of our matrix
element calculations.  At tree level in the standard model $F^W_{1L}
= 1$ and all other form factors are zero.  In fact, $F^W_{1R}=F^W_{2R}
=0$ to all orders in the standard model in the limit of massless
bottom quark.  The antitop form factors are identical to these in the
limit of CP invariance.

Similarly, the $\gamma,Z \rightarrow t\bar t $ production
vertices can be written
\begin{equation}
i{\cal M}^{i\mu} = ie
  \Bigl\lbrace
  \gamma^\mu  [F^i_{1V} + F^i_{1A}\gamma_5]
 +{i\sigma^{\mu\nu} q_\nu\over2m_t} [F^i_{2V} + F^i_{2A}
 \gamma_5]
  \Bigr\rbrace \ ,\label{prodffs}
\end{equation}
where each form factor can be a function of the center-of-mass energy
$\sqrt{s}$, the superscript is $i=\gamma,Z$, and we have again dropped
a third pair of form factors which are unobservable.  At tree level in
the standard model
$F^\gamma_{1V}={2\over3}$,
$F^Z_{1V}=({1\over4}-{2\over3}s_w^2)/s_wc_w$, and $F^Z_{1A}=
(-{1\over4})/s_wc_w$, and all others are zero.  Here, $s_w=\sin\theta_w$ and
$c_w=\cos\theta_w$.  In the limit of CP invariance $F^i_{2A}=0$.
The production analysis is simplified if we consider separately
the two possible helicities of the incoming electrons, so that the
contribution of the photon and the $Z$ add coherently.  We define new
form factors by
\begin{eqnarray}
{\cal F}^L_{ij}&=& -F^\gamma_{ij}+
\Bigl({-{1\over2}+s_w^2\over s_wc_w}\Bigr)\Bigl({s\over
s-m_Z^2}\Bigr)F^Z_{ij}\nonumber\\
{\cal F}^R_{ij} &=&-F^\gamma_{ij}+\Bigl({s_w^2\over s_wc_w}\Bigr)
\Bigl({s\over s-m_Z^2}\Bigr)
F^Z_{ij}\ ,\label{combffs}
\end{eqnarray}
where the subscripts, $i=1,2$ and $j=V,A$ refer to the structure of
the form factor, and the superscripts refer to the helicity of the
incoming electron.

We are now ready to discuss the helicity angle description of the
complete event.  As mentioned previously, in the limit of narrow width
for the top and the $W$, the event can be considered as a succession
of two-body decays.  The first process we consider is the decay of
the virtual $\gamma,Z$ boson into the $t\bar t$ pair.  Note that the
intermediate vector boson receives twice the helicity of the initial
electron, along the beam direction.  This process can be described in
the $e^+e^-$ center-of-momentum frame by two angles, the polar angle
$\theta$ and the azimuthal angle $\phi$ of the top with respect to the
electron beam axis.  Using the notation $t_L$ and $t_R$ to denote the
helicities $h_t=-1/2$ and $h_t=+1/2$, we obtain the matrix elements
\begin{eqnarray}
{\cal M}(e_L\bar e_R\rightarrow t_L\bar t_R)
&=& \bigl[{\cal F}^L_{1V}-\beta {\cal F}^L_{1A}+{\cal F}^L_{2V}\bigr]
\,(1+\cos\theta)e^{-i\phi}\nonumber\\
{\cal M}(e_L\bar e_R\rightarrow t_R\bar t_L) &=& \bigl[{\cal F}^L_{1V}+\beta
{\cal F}^L_{1A}+{\cal F}^L_{2V}\bigr]
\,(1-\cos\theta)e^{-i\phi}\nonumber\\
{\cal M}(e_L\bar e_R\rightarrow t_L\bar t_L) &=&
\gamma^{-1}\bigl[{\cal F}^L_{1V}+\gamma^2({\cal F}^L_{2V}+\beta
{\cal F}^L_{2A})\bigr]
\,(\sin\theta)e^{-i\phi}\nonumber\\
{\cal M}(e_L\bar e_R\rightarrow t_R\bar t_R) &=&
\gamma^{-1}\bigl[{\cal F}^L_{1V}+\gamma^2({\cal F}^L_{2V}-\beta
{\cal F}^L_{2A})\bigr]
\,(\sin\theta)e^{-i\phi}\nonumber\\
{\cal M}(e_R\bar e_L\rightarrow t_L\bar t_R) &=&
-\bigl[{\cal F}^R_{1V}-\beta
{\cal F}^R_{1A}+{\cal F}^R_{2V}\bigr]
\,(1-\cos\theta)e^{i\phi}                                \label{prodmat}\\
{\cal M}(e_R\bar e_L\rightarrow t_R\bar t_L) &=&
-\bigl[{\cal F}^R_{1V}+\beta
{\cal F}^R_{1A}+{\cal F}^R_{2V}\bigr]
\,(1+\cos\theta)e^{i\phi}\nonumber\\
{\cal M}(e_R\bar e_L\rightarrow t_L\bar t_L) &=&
\gamma^{-1}\bigl[{\cal F}^R_{1V}+\gamma^2({\cal F}^R_{2V}+\beta
{\cal F}^R_{2A})\bigr]
\,(\sin\theta)e^{i\phi}\nonumber\\
{\cal M}(e_R\bar e_L\rightarrow t_R\bar t_R) &=&
\gamma^{-1}\bigl[{\cal F}^R_{1V}+\gamma^2({\cal F}^R_{2V}-\beta
{\cal F}^R_{2A})\bigr]
\,(\sin\theta)e^{i\phi}\ ,\nonumber
\end{eqnarray}
where we have removed a factor of $ie^2$. Here,
$\beta^2=(1-4m_t^2/s)$ and $\gamma=\sqrt{s}/(2m_t)$.  For
longitudinally polarized beams the $\phi$ dependence will vanish.

The nice aspect of this helicity formalism is that the angular
dependence of each of the amplitudes is determined, up to a relative phase,
simply by angular momentum conservation.  For instance, in the first
matrix element the virtual vector boson has helicity -1 along the electron
beam direction, the top has helicity -1/2, and the antitop has
helicity +1/2.  To conserve angular momentum the top must move in the
electron direction and the antitop must move in the positron
direction; hence the $(1+\cos\theta)$ dependence.
By measuring the angular distributions it is straightforward to
extract the relative weights for each helicity combination, and
thereby obtain the top quark form factors.

As an example, we plot in Fig.~1 the tree-level Standard Model production
cross-section as a function of $\cos\theta$ for a top mass of 175 GeV
and a collider energy of 400 GeV for polarized electron beams.  We
have also plotted the helicity subprocesses.  Here we see that the
$e_L$'s produce predominantly $t_L$'s highly peaked in
the forward direction, while $e_R$'s produce predominantly
$t_R$'s peaked in the forward direction.  This can easily be understood
in the limit of high energy, where the $SU(2)_L\times U(1)$ symmetry
is restored and the squared matrix elements become
\begin{eqnarray}
|{\cal M}(e_L\bar e_R\rightarrow t_L\bar t_R)|^2 &=&
  \Bigl({1\over4s_w^2}+{1\over12c_w^2}\Bigr)^2
\,(1+\cos\theta)^2\ \sim\ 1.41\,(1+\cos\theta)^2\nonumber\\
|{\cal M}(e_L\bar e_R\rightarrow t_R\bar t_L)|^2 &=&
  \Bigl({1\over3c_w^2}\Bigr)^2
\,(1-\cos\theta)^2\ \sim\ 0.19\,(1-\cos\theta)^2\nonumber\\
|{\cal M}(e_R\bar e_L\rightarrow t_R\bar t_L)|^2 &=&
  \Bigl({2\over3c_w^2}\Bigr)^2
\,(1+\cos\theta)^2\ \sim\ 0.75\,(1+\cos\theta)^2      \label{highenergy}\\
|{\cal M}(e_R\bar e_L\rightarrow t_L\bar t_R)|^2 &=&
  \Bigl({1\over6c_w^2}\Bigr)^2
\,(1-\cos\theta)^2\ \sim\ 0.05\,(1-\cos\theta)^2\ ,\nonumber
\end{eqnarray}
while the remaining matrix elements vanish.  Thus, longitudinally
polarized electrons are an excellent source of polarized top quarks.

The next stage in the event is the decay of the top $t\rightarrow
bW^+$.  This process is most conveniently described in the top rest
frame obtained from the lab frame by rotating the axes $-\phi$,
then $-\theta$, and then boosting in the direction opposite to the
top momentum.  The helicity angles in this frame are
the polar angle $\chi_t$ and the azimuthal angle
$\psi_t$ of the $W$ boson with respect to the top momentum axis.
Using the notation $(L,R,Z)$ to denote the $W^+$ helicities
$(-1,+1,0)$, we obtain the
helicity amplitudes for the left-handed bottom quarks:
\begin{eqnarray}
{\cal M}(t_R\rightarrow b_LW^+_Z) &=& w^{-1}\bigl[F^W_{1L}-{1\over2}
  w^2F^W_{2L}\bigr]
  \,(\cos{\chi_t\over2})e^{i\psi_t/2}\nonumber\\
{\cal M}(t_L\rightarrow b_LW^+_Z) &=& w^{-1}\bigl[F^W_{1L}-{1\over2}
  w^2F^W_{2L}\bigr]
  \,(\sin{\chi_t\over2})e^{-i\psi_t/2}\nonumber\\
{\cal M}(t_R\rightarrow b_LW^+_L) &=& \sqrt{2}\bigl[F^W_{1L}-{1\over2}
  F^W_{2L}\bigr]\,(-\sin{\chi_t\over2})e^{i\psi_t/2}      \label{wdecaymat}\\
{\cal M}(t_L\rightarrow b_LW^+_L) &=&  \sqrt{2}\bigl[F^W_{1L}-{1\over2}
  F^W_{2L}\bigr]\,(\cos{\chi_t\over2})e^{-i\psi_t/2}\nonumber\\
{\cal M}(t_L\rightarrow b_LW^+_R) &=& {\cal M}(t_R\rightarrow
b_LW^+_R)\ =\ 0\ ,\nonumber
\end{eqnarray}
where $w=m_W/m_t$, and we have dropped an overall factor of
$igm_t(1-w^2)^{1/2}/\sqrt{2}$.
The matrix elements for right-handed bottom quarks are obtained
from these by replacing everywhere $L\leftrightarrow R$, $\psi_t
\leftrightarrow-\psi_t$, and $\chi_t\leftrightarrow-\chi_t$.

As before, the angular dependence is
exactly what is expected from angular momentum conservation in the
decay of a spin-$1/2$ object.  In addition, in the Standard Model in
the limit $m_b=0$ the top can only decay to $b_L$'s.  Therefore, it
must decay to $W^+_Z$'s in the direction of the top quark spin,
to $W^+_L$'s in the direction opposite to the top quark spin, and it
cannot decay to $W^+_R$'s at all.  In Fig.~2 we display this by
plotting the $t_R\rightarrow bW^+$ decay distribution as a function of
$\cos\chi_t$, while also plotting the helicity subprocesses.  For
increasing top mass the distribution becomes more sloped in the forward
direction, indicating an increased partial width to $W^+_Z$.

The antitop decay $\bar t\rightarrow \bar b W^-$ can be described in an
analogous manner in the antitop rest-frame,
obtained from the lab frame by rotating the axes $-\phi$,
then $\pi-\theta$, and then boosting in the direction opposite to the
antitop momentum.  The helicity angles in this frame are the polar angles
of the $W^-$, $\bar\chi_t$ and $\bar\psi_t$, with respect to the
antitop momentum axis.  If CP is a good symmetry we can obtain the matrix
elements using
\begin{equation}
{\cal M}(t_h\rightarrow b_{\rho}W^+_\lambda)\ =\
{\cal M}(\bar t_{-h}\rightarrow \bar b_{-\rho}W^-_{-\lambda})\ ,
\label{antit}
\end{equation}
while replacing $\chi_t\rightarrow\bar\chi_t$ and $\psi_t\rightarrow
-\bar\psi_t$.

The final step in the decay chain is $W^+\rightarrow\ell\nu$.
We work in the $W^+$ rest frame obtained from the top rest frame
by rotating the axes
$-\psi_t$, then $-\chi_t$, and then boosting against the $W^+$ momentum.
The helicity angles in this frame are the polar angle $\chi$
and the azimuthal angle $\psi$ of the charged lepton with respect to
the $W^+$ momentum axis.  For hadronic decays we can just replace
$\ell^+$  with the antiquark and $\nu$ with the quark.  In the
Standard model the $W^+$ can only decay to $\ell^+_R\nu_L$ in the
limit of massless leptons.  The helicity amplitudes are
\begin{eqnarray}
{\cal M}(W^+_R\rightarrow \ell^+\nu) &=&
{1\over\sqrt{2}}(1+\cos\chi)\,e^{i\psi}\nonumber\\
{\cal M}(W^+_Z\rightarrow \ell^+\nu) &=&\sin\chi     \label{wpdecay}\\
{\cal M}(W^+_L\rightarrow \ell^+\nu) &=&
{1\over\sqrt{2}}(1-\cos\chi)\,e^{-i\psi}\ ,\nonumber
\end{eqnarray}
where we have removed a factor of $igm_W/\sqrt{2}$.
In Fig.~3 we plot the $\cos\chi$ distribution in the
$W^+\rightarrow\ell^+\nu$ decay,
along with helicity subprocesses, for $W^+$ produce in top decays.
The zero at $\cos\chi=1$
indicates the absence of right-handed $W^+$'s.

Lastly,
the decay $W^-\rightarrow \ell^-\bar \nu$ can be described
in the $W^-$ rest-frame, obtained from the top rest frame
by rotating the axes
$-\bar\psi_t$, then $-\bar\chi_t$, and then boosting against the $W^-$
momentum.  The helicity angles in this frame are the polar angles
of the negatively-charged lepton, $\bar\chi$ and $\bar\psi$, with
respect to the $W^-$ momentum axis.
We can obtain the helicity amplitudes from
\begin{equation}
{\cal M}(W^+_\lambda\rightarrow \ell^+\nu)\ =\
{\cal M}(W^-_{-\lambda}\rightarrow \ell^-\bar\nu)\ ,\label{ellanti}
\end{equation}
while replacing $\chi\rightarrow\bar\chi$ and
$\psi\rightarrow-\bar\psi$.

In practice, in order to optimize the analysis of the top quark form
factors it is necessary to study the event in a
multi-dimensional space of all these angles.  The use of helicity
angles makes it easy to discern which variables are most
important for studying which form factors.  For example, by
cutting on the production angle $\theta$, while using a polarized
electron beam, it is possible to obtain a sample of highly polarized
top quarks.  With these, one can study the decay form
factors by looking at both the top decay angle $\chi_t$ and the $W$
decay angle $\chi$, in order to determine the helicities of the $W$'s.
Perhaps the optimum technique would be to use all of the helicity
angle information in a maximum likelihood fit\cite{schmidtbarklow}.
In any case we now obtain the full tree-level correlation
information of the event from
\begin{eqnarray}
&&\Bigl|\sum_{hh^\prime\lambda\lambda^\prime\rho\rho^\prime}{\cal M}(e_\sigma
\bar e_{\sigma^\prime}\rightarrow t_h\bar t_{h^\prime})
{\cal M}(t_h\rightarrow b_\rho W_\lambda^+){\cal M}(W_\lambda^+\rightarrow
\ell^+\nu)\nonumber\\&&\qquad\qquad\qquad\qquad\quad\times\
{\cal M}(\bar t_{h^\prime}\rightarrow
\bar b_{\rho^\prime}W_{\lambda^\prime}^-){\cal M}(W_{\lambda^\prime}^-
\rightarrow \ell^-\bar\nu)\Bigr|^2\label{squamp}
\end{eqnarray}
for each initial-state helicity configuration.

\section{The event at ${\cal O}(\alpha_s)$}

In the narrow top-width approximation, in which the top quarks are
treated as on-shell particles in the matrix elements, the ${\cal
O}(\alpha_s)$ corrections can be unambiguously assigned to the $t\bar
t$ production process, or to the $t$-decay or $\bar t$-decay
processes.  We have constructed a NLO Monte Carlo
by separately building a generator for $t\bar t$ events with
an extra gluon in the production, the $t$-decay, and the $\bar
t$-decay processes, as well as for events with no extra visible gluon.
To see how this is implemented it is easiest
to temporarily ignore the angular correlations.  Then the total
differential cross section, $d\sigma_{\rm tot}$,
for the event $e^+e^-\rightarrow t\bar t+X\rightarrow b\ell^+
\nu \bar b\ell^-\bar\nu+X$ is just the product of the $t\bar t+X$
differential production cross section, $d\sigma$,
times the $t$ and $\bar t$ decay distributions:
\begin{equation}
d\sigma_{\rm tot}\ =\ d\sigma\,{d\Gamma\, d\bar\Gamma
\over\Gamma^2}\ .
\label{event}
\end{equation}
To ${\cal O}(\alpha_s)$ this can be written
\begin{eqnarray}
d\sigma^{(0+1)}_{\rm tot} &=&
d\sigma^0\,{d\Gamma^0\, d\bar\Gamma^0\over(\Gamma^0)^2}
\ +\ d\sigma^1\,{d\Gamma^0\, d\bar\Gamma^0
\over(\Gamma^0)^2}
\ +\ d\sigma^0\,{d\Gamma^1\, d\bar\Gamma^0
\over(\Gamma^0)^2}
\nonumber\\
&&
+\ d\sigma^0\,{d\Gamma^0\, d\bar\Gamma^1
\over(\Gamma^0)^2}
\ -\ {2\Gamma^1\over\Gamma^0}\,
d\sigma^0\,{d\Gamma^0\, d\bar\Gamma^0\over(\Gamma^0)^2}
\ ,\label{event1}
\end{eqnarray}
where the first term is the tree-level event, the second term includes
${\cal O}(\alpha_s)$ corrections to the $t\bar t$ production, the
third and fourth terms contain the corrections to the $t$ and the
$\bar t$ decay respectively, and the last term is the ${\cal
O}(\alpha_s)$ correction to the widths in the denominator.  Note that
on integrating over the decay phase space, the last three terms cancel so
that $\sigma^{(0+1)}_{\rm tot}=\sigma^0+\sigma^1\,$,
{\it i.e.}, the integrated total event cross section is not affected
by the corrections to the top quark decay, as required.

The ${\cal O}(\alpha_s)$ corrections to the production and decay can
be separated into three pieces---the virtual (v), soft-gluon (s), and
real-gluon (r) contributions:
\begin{eqnarray}
d\sigma^1 &=&
d\sigma^{v}\ +\  d\sigma^{s}(x_0)\ +\
d\sigma^{r}(x_0) \nonumber\\
d\Gamma^1 &=& d\Gamma^{v} \ +\ d\Gamma^{s}(y_0,z_0) \ +\
d\Gamma^{r}(y_0,z_0) \
.\label{separation}
\end{eqnarray}
The arbitrary distinction between ``soft'' and ``real'' gluons is implemented
using artifical cutoffs $x_0, y_0, z_0$, which we will describe more fully
below.  The real gluons are defined to be those produced above the
cutoffs and are treated using the exact 3-body phase space.
The soft gluons are those produced below the
cutoffs and are integrated out analytically, leaving an effective 2-body
phase space.  Both the virtual and the soft contributions are infrared
divergent, but their sum is infrared finite.
Thus, we can combine the virtual and soft contributions,
and we can conveniently separate the full ${\cal O}(\alpha_s)$ cross
section into the sum of four sub-event cross sections:
\begin{eqnarray}
d\sigma^{(0+1)}_{\rm tot} &=&
d\sigma^{(v+s)}_{\rm tot}(x_0,y_0,z_0)\ +\
d\sigma^{r}(x_0)\,{d\Gamma^0\, d\bar\Gamma^0
\over(\Gamma^0)^2} \nonumber\\
&&
+\ d\sigma^0\,{d\Gamma^{r}(y_0,z_0)\, d\bar\Gamma^0
\over(\Gamma^0)^2}
\ +\ d\sigma^0\,{d\Gamma^0\, d\bar\Gamma^{r}
(y_0,z_0)\over(\Gamma^0)^2}
\ .\label{subevents}
\end{eqnarray}
The last three contributions have 7 final-state partons, containing a
real gluon in the production, the $t$ decay, or the $\bar t$ decay
respectively.  Each of these terms is manifestly positive-definite.
The first contribution has only 6 final-state
partons and is given by the following sum:
\begin{eqnarray}
d\sigma^{(v+s)}_{\rm tot}(x_0,y_0,z_0) &=&
\Bigl(1-{2\Gamma^1\over\Gamma^0}\Bigr)\,
d\sigma^0\,{d\Gamma^0\, d\bar\Gamma^0
\over(\Gamma^0)^2}
\ +\ d\sigma^{(v+s)}(x_0)\,{d\Gamma^0\, d\bar\Gamma^0
\over(\Gamma^0)^2}                             \label{virtualglue}\\
&&
+\ d\sigma^0\,{d\Gamma^{(v+s)}(y_0,z_0)\, d\bar\Gamma^0
\over(\Gamma^0)^2}
\ +\ d\sigma^0\,{d\Gamma^0\, d\bar\Gamma^{(v+s)}
(y_0,z_0)\over(\Gamma^0)^2}
\ .\nonumber
\end{eqnarray}
This term may be negative for small values of the cutoffs.
A separate Monte Carlo is used to generate events for each of the four
terms in (\ref{subevents}) with all angular correlations included.

We now elaborate on the infrared cancellations, as well as the
separation into ``soft'' and ``real'' gluons, that are used in equation
(\ref{subevents}).  The virtual corrections to the production and
decay processes can be written as corrections to the form factors
(\ref{decayffs}) and (\ref{prodffs}), with the understanding that they
are only expanded to ${\cal O}(\alpha_s)$ in the squared amplitudes
(\ref{squamp}).  Using dimensional
regularization with $D=4-2\epsilon$ we obtain the correction to the
production form factors:
\begin{eqnarray}
\delta F^i_{1V}&=& {\alpha_sC_q\over2\pi}\,f^i_{V}\,(I_1+I_2)\nonumber\\
\delta F^i_{1A}&=& {\alpha_sC_q\over2\pi}\,f^i_{A}\,(I_1-I_2)
\label{prodcorr}\\
\delta F^i_{2V}&=& {\alpha_sC_q\over2\pi}\,f^i_{V}\,(2I_2)\nonumber
\end{eqnarray}
where
\begin{eqnarray}
I_1&=&\biggl({4\pi\mu^2\over m_t^2}\biggr)^\epsilon\Gamma(1+\epsilon)
\,\Biggl\{{1\over\epsilon}\,\biggl[-1-{1+\beta^2\over2\beta}\,\Bigl(
\ln{1-\beta\over1+\beta}+i\pi\Bigr)\biggr]\
-\,2 \nonumber\\
&&\qquad\qquad\qquad\quad\,
+\,{1+\beta^2\over2\beta}\,\biggl[\Bigl(-{3\over2}+\ln
{4\beta^2\over1-\beta^2}\Bigr)\,\Bigl(
\ln{1-\beta\over1+\beta}+i\pi\Bigr)
\nonumber\\ &&\qquad\qquad\qquad\qquad\qquad\quad + {2\pi^2\over3} +
2{\rm Li}_2\Bigl({1-\beta\over1+\beta}\Bigr) + {1\over2}\Bigl(
\ln{1-\beta\over1+\beta}\Bigr)^2\biggr]\Biggr\}
\nonumber\\
I_2&=&{1-\beta^2\over4\beta}\,\Bigl[
\ln{1-\beta\over1+\beta}+i\pi\Bigr]\ ,
\label{bigis}
\end{eqnarray}
$C_q=4/3$, and
$f^\gamma_{V}={2\over3}$, $f^\gamma_{A}=0$,
$f^Z_{V}=({1\over4}-{2\over3}s_w^2)/s_wc_w$, and $f^Z_{A}=
(-{1\over4})/s_wc_w$ are the tree-level couplings.  Also, we have used the
Spence function ${\rm Li}_2(z)=-\int_0^zdt\ln(1-t)/t$.
This agrees with the
previous results given in Ref.~\cite{JLZ}.  Note that the
contribution from ${\rm Re}{I_1}$ is proportional to the tree-level cross
section, while ${\rm Im}{I_1}$ does not contribute at ${\cal
O}(\alpha_s)$.

For the real gluon corrections to $t\bar t$ production it is
convenient to define the gluon phase space in terms of the variables
\begin{equation}
x\ =\ E_g/E^{\rm max}_g\ \ ,\qquad\quad \Delta\ =\
(1-\cos\theta^*_{tg})/2\ \ ,\label{prodcut}
\end{equation}
where the maximum energy of the production gluon in the lab frame is
$E^{\rm max}_g=\beta^2\sqrt{s}/2$, and $\theta^*_{tg}$ is the
angle between the gluon and top quark momenta in the $t\bar t$ rest
frame.  The full phase space is $0<x<1$, $0<\Delta<1$ with the soft
gluon limit given by $x\rightarrow0$.  Integrating out the
gluons in the region $x<x_0$, for small $x_0$, it is possible to
absorb this soft-gluon contribution into the form factors
(\ref{prodcorr}) by replacing $I_1\rightarrow I_1+I^{\rm (soft)}_1$,
with
\begin{eqnarray}
I^{\rm (soft)}_1&=&\biggl({4\pi\mu^2\over m_t^2}\biggr)^\epsilon
\Gamma(1+\epsilon)\,
\Biggl\{\Bigl({1\over\epsilon}-2\ln{x_0}\Bigr)\,
\biggl[1+{1+\beta^2\over2\beta}\,
\ln{1-\beta\over1+\beta}\biggr]\nonumber\\
&&\qquad\qquad\qquad\quad
+\ln{1-\beta^2\over4\beta^4}\, -\,{1\over\beta}
\ln{1-\beta\over1+\beta}\nonumber\\
&&\qquad\qquad\qquad\quad
+\, {1+\beta^2\over2\beta}\,\biggl[
-\ln\beta^2\ln{1-\beta\over1+\beta}
-{\pi^2\over3}
                                                 \label{bigisoft}\\
&&\qquad\qquad\qquad\qquad\qquad\quad
+2{\rm Li}_2\Bigl({1-\beta\over1+\beta}\Bigr) + {1\over2}\Bigl(
\ln{1-\beta\over1+\beta}\Bigr)^2\biggr]\Biggr\}
\ .
\nonumber
\end{eqnarray}
The sum of the virtual and soft contributions ${\rm Re}(I_1+I^{\rm
(soft)}_1)$ is
now IR finite.  The ``real'' gluons with $x>x_0$ are treated using
exact kinematics.  The matrix elements can be written in terms of
helicity amplitudes as in section 2.  We leave the details of this to
the appendix A.

The virtual corrections to the top decay form factors at ${\cal
O}(\alpha_s)$ are
\begin{eqnarray}
\delta F^W_{1L}&=& {\alpha_sC_q\over2\pi}\,
\biggl({4\pi\mu^2\over m_t^2}\biggr)^\epsilon\Gamma(1+\epsilon)\,
\Biggl\{-{1\over2\epsilon^2}+{1\over\epsilon}\,
\Bigl[-{5\over4} +\ln{(1-w^2)}\Bigr]\nonumber\\
&&\qquad\qquad\qquad\qquad-3-\Bigl(\ln{(1-w^2)}\Bigr)^2\nonumber\\
&&\qquad\qquad\qquad\qquad
+{3\over2}\ln{(1-w^2)}-{\rm Li}_2(w^2)\Biggr\}
\label{decaycorr}\\
\delta F^W_{2L}&=& {\alpha_sC_q\over2\pi}w^{-2}\ln{(1-w^2)}\ .\nonumber
\end{eqnarray}
For the phase space of the real gluon in top decay
we use the variables
\begin{equation}
y\ =\ E_g/E^{\rm max}_g\ \ ,\qquad\quad z\ =\
(1-\cos\theta^*_{bg})/2\ ,\label{decaycut}
\end{equation}
where the maximum energy of the decay gluon in the top rest frame is
$E^{\rm max}_g=(m_t/2)(1-w^2)$, and $\theta^*_{bg}$ is the
angle between the gluon and bottom quark momenta in the $bW$ rest
frame.  The gluon becomes soft in the limit $y\rightarrow0$
and collinear in the limit $z\rightarrow0$.
Integrating out the soft and collinear gluons for which $y<y_0$ and/or
$z<z_0$, for small $y_0,z_0$, we can absorb these contributions into
the form factor $F^W_{1L}$.  They contribute
\begin{eqnarray}
\delta F^{W{\rm (soft)}}_{1L}&=& {\alpha_sC_q\over2\pi}\,
\biggl({4\pi\mu^2\over m_t^2}\biggr)^\epsilon\Gamma(1+\epsilon)\,
\Biggl\{{1\over2\epsilon^2}+{1\over\epsilon}\,
\Bigl[\,{5\over4} -\ln{(1-w^2)}\Bigr]\nonumber\\
&&\qquad\qquad +4 +{5-3w^2\over8(1-w^2)}
+\Bigl(\ln{(1-w^2)}\Bigr)^2\nonumber\\
&&\qquad\qquad
-{5\over2}\ln{(1-w^2)}+{\rm Li}_2(1-w^2)          \label{decaysoft}\\
&&\qquad\qquad
-{w^2(2-3w^2)\over4(1-w^2)^2}\,\ln{w^2}-{\pi^2\over2}
\nonumber\\ &&\qquad\qquad
-(1+\ln{x_0})(1+\ln{z_0})\,+\,{1\over4}\ln{z_0}
\Biggr\}\ ,\nonumber
\end{eqnarray}
so that the sum of the virtual and soft-gluon contributions $\delta
F^W_{1L}+ \delta F^{W{\rm (soft)}}_{1L}$ is
IR finite.  As in the production process, the ``real'' gluons with
$y>y_0$ and $z>z_0$ are treated using
exact kinematics.  The helicity amplitudes are given in the appendix B.

It is useful at this stage to describe the Monte Carlo more fully.
It is written in the C++ programming language and contains a separate
event-generator class for each of the four sub-channel processes in
equation (\ref{subevents}).  Each of these sub-channel generators are
in turn derived
from a single tree-level generator which produces the helicity
angles of the event with the exact tree-level distributions.  The
sub-channel generators then produce the relevant gluon kinematic
variables, prepare the particle four-vectors, and give the event a
weight.  The production-gluon class generates the gluon variables
(\ref{prodcut}) with a soft gluon distribution, while the decay-gluon
classes generate the gluon variables (\ref{decaycut}) with a soft and
collinear gluon distribution.  This results in a very efficient Monte
Carlo for each of the four sub-channels.

The relative contribution from the four sub-channels depends on the
artificial IR
cutoffs ($x_0,y_0,z_0$).  The choice of values for these parameters is
determined by several considerations.  First, the analytic integrations
of the soft gluons contained in (\ref{bigisoft}) and (\ref{decaysoft})
are valid up to terms linear in the cutoffs, so they should be
kept as small as possible.  In addition, they
should lie below any physical cutoff, determined by the detector energy
resolution or the jet definition.  However, for very small cutoffs the
contribution containing the virtual and soft gluons will become very
large and negative, and there will be large cancellations between it and the
other sub-channels.  Thus, the cutoffs should not be too small or else
the numerical errors will become prohibitive.  Luckily, this last
constraint turns out to be not too restrictive for our Monte Carlo.
For each plot in the next two sections we have checked that the results
do not change significantly for smaller values of the cutoffs.
As a final test of our confidence,
we have checked that our Monte Carlo reproduces the
${\cal O}(\alpha_s)$ production\cite{JLZ} and
decay distributions\cite{JezKuhn} of previous analyses.

Our Monte Carlo also allows the inclusion of width effects by
generating Breit-Wigner resonance distributions for the tops and the
$W$'s.  In addition, the kinematic effects of the
bottom quark mass can be included.  Momentum conservation is maintained
by shifting the energies of the final-state particles, while keeping
the helicity angles and the gluon kinematic variables (\ref{prodcut})
and (\ref{decaycut}) fixed.  This procedure should be good to
${\cal O}(\Gamma_t/m_t)$ except very near threshold.  Note, however,
that the matrix elements, and hence the event weights, are always
computed in the zero-width and $m_b=0$ limits.
Finally, initial state radiation (ISR) can be included by generating
electron and positron momentum fractions $z$ with the
distribution function given by Fadin and Kuraev\cite{FK}
\begin{equation}
D_e(z)\ =\ \hat\beta/2(1- z)^{\hat\beta/2-1}\,
(1+3\hat\beta/8) -\hat\beta(1+ z)/4\ ,
\label{electdist}
\end{equation}
where $\hat\beta=(2\alpha/\pi)(\ln{s/m_e^2}-1)$.

It must be noted that the narrow-width approximation is necessary for
the NLO analysis of this section.  As a consequence,
the Monte Carlo does not include the effects of
interference between gluons emitted in the production and gluons
emitted in the decay.  These perturbative effects have been studied in
the soft-gluon limit in Refs.~\cite{softinterference}.
Typically, the interference is only important for gluons with energy
$E_g\le\Gamma_t$.  However, it should be considered in a complete analysis.
In addition, because the final-state bottom quarks do carry bare color,
there will be some nonperturbative information connecting them in
the form of soft hadrons\cite{nonpert}.  We have neglected this
effect here.

\section{The effects of radiated gluons}

In this section we will study the top quark mass reconstruction and
helicity angle distributions at next-to-leading order.  We do
this by starting with an ideal event situation---no ISR, an ideal
$4\pi$ detector, perfect partonic-level particle identification.  In
the subsequent section we will make each of these factors more
realistic experimentally.  The purpose here is to develop our
intuition by isolating the purely theoretical QCD effects at NLO.
If we assume that both bottom quarks and $W$'s are identified and
signed and that there is $4\pi$ detector coverage, then the only
ambiguity is in where to put the gluon.  Does it belong to the $t$,
to the $\bar t$, or to neither?

Here, we will make this assignment of the real gluon in analogy with
the typical jet-clustering algorithm used at $e^+e^-$ colliders.  Defining
the quantities $\mu^2=(p_b+p_g)^2$ and $\bar\mu^2=(\bar
p_b+p_g)^2$, we make the assignment:
\renewcommand{\arraystretch}{1.5}
\begin{equation}
\begin{array}{lcl}
\mbox{if $\mu<\bar\mu$ and $\mu<\mu_{cut}$} &\quad\Rightarrow\quad&
\mbox{gluon belongs to $t$ decay}\\
&&(p_t=p_g+p_b+p_{W^+}\ , \ \bar p_t=\bar p_b+p_{W^-})\ ,  \\
\mbox{if $\bar\mu<\mu$ and $\bar\mu<\mu_{cut}$} &\quad\Rightarrow\quad&
\mbox{gluon belongs to $\bar t$ decay} \\
&&(p_t=p_b+p_{W^+}\ , \ \bar p_t=p_g+\bar p_b+p_{W^-})\ , \\
\mbox{else} &\quad\Rightarrow\quad&\mbox{gluon belongs to
production} \\
&&(p_t=p_b+p_{W^+}\ , \ \bar p_t=\bar p_b+p_{W^-})\ .
\end{array}\label{mu}
\end{equation}
In the limit $m_b=0$, we recognize $\mu_{cut}$ as an infrared cutoff
on both the collinear and soft gluons in the event.  In fact,
we can consider the decay gluons to be clustered with the
bottom quarks\cite{cluster} using the standard jet resolution
parameter $y_{cut}=\mu_{cut}^2/s$.  By varying
$\mu_{cut}$ we change the fraction of events with gluons that are not
combined with the $b$ or $\bar b$, and thus are considered to be part
of the production process.  This fraction is plotted in Fig.~4 as a
function of the center-of-mass energy for various values of $\mu_{cut}$.
At this fixed order in perturbation theory, the fraction can be
greater than one, indicating that a resummation of the large
logarithms in $y_{cut}$ or $\mu_{cut}^2/m_t^2$ is necessary.
As in all of our plots we use a standard top mass of 175 GeV and
$\alpha_s=0.12$.

We now consider the top quark mass distribution at $\sqrt{s}=400$
GeV.  Using the algorithm (\ref{mu}), each event produces two mass
values $m^2=p_t^2$ and $\bar m^2=\bar p_t^2$, which
are binned independently.  To see most
clearly how the radiation affects this distribution, we plot it in
Fig.~5(a) in the strict zero top-width limit for values of
$\mu_{cut}=5$, 10, 20,
and $\infty$ GeV.  Note that for $\mu_{cut}=\infty$, all of the observed
gluons are assigned either to top decay or antitop decay, and none to
the production.  The Monte Carlo cutoffs used are
$x_0=0.02$, $y_0=0.005$, and $z_0=0.01$.  The $\delta$-function
spike in the central bin arises from those events in which the top
momentum is determined correctly from its true decay products.  The
excess below the $\delta$-function corresponds to events where
a decay gluon is assigned incorrectly and is not included in the top
momentum reconstruction.  These
missed-gluon events become less likely as $\mu_{cut}$ increases, but even
for $\mu_{cut}=\infty$ there is a remnant of events where the gluon gets
assigned to the wrong-charge top quark.  The excess above the
$\delta$-function
corresponds to events where an extra gluon is incorrectly included in
a top momentum reconstruction.  This region has two separate
contributions, from mis-assigned decay gluons and from mis-assigned
production gluons.  Both of these increase with increasing $\mu_{cut}$,
with the production gluons adding a second bump for larger values of
this parameter.
The deficits in the distribution directly on each side of the spike are
due to the artificial cutoffs $x_0$, $y_0$, and $z_0$.

The $\delta$-function peak in this distribution is an artifact of the
zero-width approximation.  Turning on the
Breit-Wigner resonance for the top quark effectively smears over the
$\delta$-function and results in a well-defined IR-finite
mass distribution.  In Fig.~5(b) we plot this distribution using the
same values of $\mu_{cut}$ as before.  For comparison, we also plot the
initial Breit-Wigner distribution.  We now choose the
Monte Carlo cutoffs to be
$x_0=0.002$, $y_0=0.0013$, and $z_0=0.0028$.  These cutoffs ensure
that all production gluons with $E_g>100$ MeV and all decay gluons
with $\mu,\bar\mu>5$ GeV are treated with exact kinematics.
The distributions do not change significantly for smaller
$x_0,y_0,z_0$.  For
$\mu_{cut}=5$ GeV we see that the mass distribution is severely
distorted, while for higher values of $\mu_{cut}$ it quickly regains
an approximate
Breit-Wigner shape, with a small decrease in the peak and an increase
in the tail regions.  We cannot take the $\mu_{cut}=5$
GeV curve too seriously, however, because for small values of $\mu_{cut}$
we are probing the collinear-gluon region of the decay phase space.
On the other hand, the effects of soft-gluon singularities are
inconsequential, because
soft gluons have $E_g\approx 0$ and do not affect the mass
measurement.  For $\mu_{cut}\gsim20$ GeV these perturbative mass
distributions should be reliable.  Fig.~5(b) suggests that perhaps the
best approach to mass reconstruction at $\sqrt{s}=400$ GeV is to treat
each extra gluon as coming from decay, combining it with whichever
top quark has the smaller value of $\mu$.  This is because 400 GeV is
still not too far from threshold, where real gluon radiation in the
production process is suppressed.

At higher energies the situation changes dramatically.
In Fig.~6 we plot the mass
distributions at $\sqrt{s}=1$ TeV for $\mu_{cut}=5$, 20, 80 and
$\infty$ GeV.  At this center-of-mass energy we choose $x_0=.0001$ so
that production gluons
with $E_g>100$ Mev are treated with exact kinematics.
The best resonant peak occurs for
$\mu_{cut}\sim20$ GeV.  At this high energy there is substantial
collinear radiation in the $t\bar t$ production process, so that for
larger values of $\mu_{cut}$ an extra gluon is usually included with
one of the tops, resulting in a too-large mass reconstruction.
These curves are suggestive of the degradation that will occur at
this energy, but a resummation of the collinear gluons would be
necessary to obtain an exact prediction.  Certainly, determining
the top mass at $\sqrt{s}=1$ TeV would be more difficult than at lower
energies.

We now turn to the top production angle distribution.  For the
remainder of this section, we work in the strict zero-width and
$m_b=0$ limits.  The production angle distribution has
been studied before at ${\cal O}(\alpha_s)$ for the pure $t\bar t$
production process in \cite{JLZ}.  Here we include the effects of
radiative corrections in both production and decay of the top quarks.
Although the corrections to the decay process do not affect
this distribution for perfectly reconstructed $t\bar t$
events, they are significant when reconstruction ambiguities are
considered.  For a given value of
$\mu_{cut}$ we can use the algorithm (\ref{mu}) to reconstruct each
event and then bin with respect to the top and
antitop production variables $\cos\theta$ and $-\cos\bar\theta$.
The tree-level production angle distributions for $m_t=175$ GeV and
$\sqrt{s}=400$ GeV were shown in Fig.~1.  In Fig.~7 we plot the
deviations from the tree-level distribution for several different
values of $\mu_{cut}$ for left- and right-handed electron beams.  We also
plot the pure production corrections\cite{JLZ}, which assume perfect
gluon discrimination and event reconstruction.  For both electron
polarizations the ${\cal O}(\alpha_s)$ corrections tend to increase the
slope of the distribution with production angle.  However, the
treatment of the radiative gluon can have a significant effect on this
correction.   For a left-polarized electron beam, using smaller values
of $\mu_{cut}$, the correction even changes sign.  This is shown
further in Table 1, where we give the ${\cal O}(\alpha_s)$ corrections
to the forward-backward asymmetry of the top quarks for the different
values of $\mu_{cut}$.

In Fig.~8 we examine the effects of the gluon ambiguity on the decay
angle of the top quark to the $W^+$ boson, $\chi_t$.  Using the
algorithm (\ref{mu}) the $W^+$ boson is reconstructed
correctly, but the observed momentum of the top quark, and therefore
the observed value of $\chi_t$, is affected by the treatment of
the radiative gluon.  In Fig.~8 we plot the fraction of observed
values of $\cos\chi_t$ falling in each 0.1-width bin for events with true
values of $\cos\chi_t$ between -0.1 and 0.0.  For small
$\mu_{cut}$ the reconstructed values of  $\cos\chi_t$ tend to be
larger than the true values.  The
missed gluons in the decay lead to an underestimate of the top
momentum, which results in an underestimate of the angle between the
$W^+$ and the top momenta after boosting to the top rest frame.  As in
the previous examples,
the most accurate reconstruction occurs for large $\mu_{cut}$.

\section{More detailed analysis of top mass reconstruction}

In this section we re-examine the top mass distribution with
more realistic experimental assumptions.  The neutrinos are
undetected and the quark jets are indistinguishable.
We include the effects of initial-state radiation, and we
impose simple lab-frame angular cuts to approximate the effects of
the detector.  We also examine the effects of parton energy smearing
due to the detector resolution.
However, we stop short of including final-state hadronization.  This
analysis is strictly at the partonic level.

We will consider the reconstruction of the top quarks in both the
lepton+jets mode and the all-jets mode.  We require that all of
the visible partons must satisfy $|\cos{\theta_{lab}}|<0.9$, and we
cluster\cite{cluster} the colored partons into jets using the
jet resolution parameter
$y_{cut}=\mu_{cut}^2/s$ with $\mu_{cut}=20$ GeV.  We do not
consider the effects of $b$-tagging, treating all hadronic jets as
indistinguishable.  We then use a simple algorithm for $t\bar t$ event
reconstruction in each mode.  Certainly, these methods can be improved
and optimized, but they will be sufficient for our purposes.

In the all-jets mode we require that there be $\ge$ 6 jets after the
cuts and the clustering.  We then choose two pairs of jets to form the
$W$'s by minimizing the quantity
\begin{equation}
\Bigl[(p_1+p_2)^2-m_W^2\Bigr]^2+\Bigl[(p_3+p_4)^2-m_W^2\Bigr]^2
\label{minimize}
\end{equation}
over all combinations of jets.  We then combine one or more of the
remaining jets with each of the $W$'s, so as to minimize the mass
difference between the resulting top quarks.

In the lepton+jets mode we require that there be a charged lepton
and $\ge$ 4 jets after the cuts and clustering.  The neutrino
four-momentum is defined to be equal to the missing momentum in the
event, $p_\nu=p_{total}-\sum p_{visible}$, with the additional
requirement that
\begin{equation}
|m(\ell\nu)-m_W|<10\ {\rm GeV}\ .\label{nconstraint}
\end{equation}
Then a pair of jets is
chosen to form the second $W$ boson by minimizing
$|(p_1+p_2)^2-m_W^2|$ over all of the jets.  Finally, we combine at
least one of the remaining jets with each of the $W$'s, so as to
minimize the mass difference between the resulting top quarks.

We begin our study by including the initial-state radiation, but
omitting the final-state energy smearing.
The mass distributions for the all-jet channel
and for the lepton+jets channel are shown in Figs.~9(a) and 9(b),
respectively, for $m_t=175$ GeV and $\sqrt{s}=400$ GeV.  For comparison
we also show the original Breit-Wigner distributions, as well as the
mass reconstructions at tree-level for both channels.  The ${\cal
O}(\alpha_s)$ distributions exhibit a moderate degradation as compared
to tree-level and also as compared to the $\mu_{cut}=\infty$ curve of
Fig.~5(b) from the previous section.  This is due to the additional
complexity in clustering the radiated gluon and reconstructing the
event.  Naturally, these effects are more serious in
the all-jet channel.  In the
lepton+jets channel there can also be errors in the neutrino
reconstruction due to initial-state radiation.  This is the
source of the enhanced tail at higher masses.
Of the all-jet events, 41\% survive the cuts and are identified as a
6-jet event, while  4\% are identified with 7 jets. Of the lepton+jet
events, 35\% survive the cuts and are identified with
4 jets, while  7\% are identified with 5 jets.

In Fig.~10 we show the same distributions with the final-state partons
smeared in energy to approximate the effects of the detector energy
resolution.  The hadronic and leptonic final-state partons are
gaussian-smeared with the parameters used in the JLC study\cite{JLC}:
\begin{equation}
{\sigma_E^{had}\over E}\ =\ {0.4\over\sqrt{E}}\ ,\qquad
{\sigma_E^{lep}\over E}\ =\ {0.15\over\sqrt{E}}\ ,
\label{smearing}
\end{equation}
where $E$ is in GeV.  The smearing has no effect on the efficiency
in the all-jets mode, but it does reduce the efficiencies in the
lepton+jet modes to 22\% (4 jets) and 4\% (5 jets).  This is because,
when the jet energies are smeared, the reconstructed neutrino is less
likely to meet the constraint
(\ref{nconstraint}).  From Fig.~10 we conclude
that the major contribution to the error on the top
mass distribution will probably come from the detector energy
resolution, making a direct width measurement virtually impossible.
The gluon radiation also contributes a significant amount to
the widening of the peak, especially in the all-jets reconstruction
channel.  As we have shown in this paper, this QCD radiative
contribution is directly calculable in perturbation theory.

The plots in this section are representative of the accuracy that may
be obtainable in a direct mass measurement, although certainly the
reconstruction algorithm can be better optimized, and $b$-tagging
would be very useful in this regard.  As for the angular
distributions, we would expect the detector resolution effects to be
less serious because detector angular resolution is usually better
than energy resolution.  However, the reconstruction errors may still
be significant for these distributions.

\section{Conclusions}

As in any hard scattering process, the $e^+e^-\rightarrow t\bar t$
event is certainly more complex than the basic tree-level parton cross
section would indicate.  The first step to a more realistic treatment
should include QCD radiation in the final state.  This requires the
correct handling of radiation both in the $\gamma^*\rightarrow t\bar
t$ production process and in the $t\rightarrow bW^+$ decay process.
In this paper we have shown how to include this radiation to ${\cal
O}(\alpha_s)$ and have constructed a Monte Carlo generator to study
these effects.  In doing this we have made strong use of the helicity
angle formalism, which is the most natural for investigating the
properties of the top quark.

The treatment of the $t\bar t$ event at ${\cal O}(\alpha_s)$
introduces reconstruction ambiguities whenever there is real gluon
radiation.  We have shown how this can alter the top mass distribution
and the angular distributions.  By including the Breit-Wigner
resonance shape for the top quark, we obtain an infrared finite
correction to the mass distribution.  The major effect of the QCD
radiation is to degrade the peak, with practically no shift in the
position of the maximum.  For energies not too far above the $t\bar t$
threshold, most of the gluon radiation occurs during the decay of the
quarks; however, at higher energies the radiation off the tops during
the production phase becomes more important.

\section*{Acknowledgements}

I wish to thank Michael Peskin for collaboration in the early stages
of this project.  I would also like to thank him and Timothy Barklow
for many useful suggestions and comments.

\section*{Appendices}

\appendix

\section{$e^+e^-\rightarrow t\bar tg$ production amplitudes}

The real radiative corrections to $t\bar t$ production and decay can
be given by helicity amplitudes, with only minor complications due to
the three-body final state.  We can describe the $t\bar t g$
production event
configuration in the lab frame in terms of five variables.  Two of
these are the energy fractions $x_i=2E_i/\sqrt{s}$ of
the top and the gluon, which are in turn determined by the variables
of
equation (\ref{prodcut}).  These fix all of the lab-frame energies and
angles within the $t\bar tg$-plane.  Two more variables are just the
polar angle $\theta$ and azimuthal angle $\phi$ of the top quark with
respect to the electron beam axis.  The final variable that we need is
the angle $\phi_g$ between the $e^+e^-t$-plane and the $t\bar
tg$-plane, rotated around the top momentum axis.  Note that the
rotation by $\phi_g$ around the top quark momentum axis also rotates
its decay products.
This completely determines the event kinematics.

For longitudinally polarized electrons, the intermediate photon-$Z$
state will be an eigenstate of spin along the beam axis.  However,
it is more convenient to work in a basis where the vector boson is a
spin eigenstate along the top momentum direction.  Labeling
these eigenstates by $\gamma_\lambda$, we can expand the matrix
elements in terms of amplitudes in the new basis, which are now
independent of the variables $\phi$, $\theta$, and $\phi_g$:
\begin{eqnarray}
\lefteqn{{\cal M}(e_L\bar e_R\rightarrow t\bar tg)
\ =\ e^{-i\phi}\biggl\{} \hspace{.3in}  \nonumber\\
&&\phantom{+} \Bigl[
{\cal F}^L_{1L}{\cal M}(L;\gamma_L\rightarrow t\bar tg)
+{\cal F}^L_{1R}{\cal M}(R;\gamma_L\rightarrow t\bar tg)
\Bigr]\,{1\over\sqrt{2}}(1+\cos\theta)e^{-i\phi_g}\nonumber\\
&&+\Bigl[
{\cal F}^L_{1L}{\cal M}(L;\gamma_R\rightarrow t\bar tg)
+{\cal F}^L_{1R}{\cal M}(R;\gamma_R\rightarrow t\bar tg)
\Bigr]\,{1\over\sqrt{2}}(1-\cos\theta)e^{i\phi_g}\nonumber\\
&&+\Bigl[
{\cal F}^L_{1L}{\cal M}(L;\gamma_Z\rightarrow t\bar tg)
+{\cal F}^L_{1R}{\cal M}(R;\gamma_Z\rightarrow t\bar tg)
\Bigr]\,\sin{\theta}\biggr\}\nonumber\\
\lefteqn{{\cal M}(e_R\bar e_L\rightarrow t\bar tg)
\ =\ e^{i\phi}\biggl\{}\hspace{.3in}              \label{zzx} \\
&&-\Bigl[
{\cal F}^R_{1L}{\cal M}(L;\gamma_L\rightarrow t\bar tg)
+{\cal F}^R_{1R}{\cal M}(R;\gamma_L\rightarrow t\bar tg)
\Bigr]\,{1\over\sqrt{2}}(1-\cos\theta)e^{-i\phi_g}\nonumber\\
&&-\Bigl[
{\cal F}^R_{1L}{\cal M}(L;\gamma_R\rightarrow t\bar tg)
+{\cal F}^R_{1R}{\cal M}(R;\gamma_R\rightarrow t\bar tg)
\Bigr]\,{1\over\sqrt{2}}(1+\cos\theta)e^{i\phi_g}\nonumber\\
&&+\Bigl[
{\cal F}^R_{1L}{\cal M}(L;\gamma_Z\rightarrow t\bar tg)
+{\cal F}^R_{1R}{\cal M}(R;\gamma_Z\rightarrow t\bar tg)
\Bigr]\,\sin{\theta}\biggr\}\ .\nonumber
\end{eqnarray}
We have also separated the pieces arising from the left-handed and
right-handed currents. The form factors
${\cal F}^i_{1R}={\cal F}^i_{1V}+{\cal F}^i_{1A}$ and
${\cal F}^i_{1L}={\cal F}^i_{1V}-{\cal F}^i_{1A}$ are
obtained from equation (\ref{combffs}) evaluated at tree level.

The matrix elements in equation (\ref{zzx}) with left-handed currents are:
\begin{eqnarray}
{\cal M}(L;\gamma_L\rightarrow t_L\bar t_Lg_L)&=&
-A_{+-}
\sin{\theta_{tg}\over2}\cos{\theta_{tg}\over2}\cos{\theta_{t\bar t}\over2}
\bigl(x_t\beta_t+(1-x_t)\bigr)\nonumber\\
{\cal M}(L;\gamma_R\rightarrow t_L\bar t_Lg_L)&=&
\phantom{-}A_{+-}
\sin^2{\theta_{tg}\over2}\sin{\theta_{t\bar t}\over2}(1-x_t)
\nonumber\\
{\cal M}(L;\gamma_Z\rightarrow t_L\bar t_Lg_L)&=&
-{A_{+-}\over\sqrt{2}}
\sin{\theta_{tg}\over2}
\bigl(x_t\beta_t\cos{\theta_{tg}\over2}\sin{\theta_{t\bar t}\over2}
+(1-x_t)\sin{\theta_{t\bar t}-\theta_{tg}\over2}\bigr)\nonumber\\
{\cal M}(L;\gamma_L\rightarrow t_L\bar t_Lg_R)&=&
\phantom{-}A_{+-}\sin{\theta_{tg}\over2}
\bigl(x_t\beta_t\cos{\theta_{tg}\over2}\cos{\theta_{t\bar t}\over2}
+(1-\bar x_t)\cos{\theta_{t\bar t}+\theta_{tg}\over2}\bigr)\nonumber\\
{\cal M}(L;\gamma_R\rightarrow t_L\bar t_Lg_R)&=&
\phantom{-}0\nonumber\\
{\cal M}(L;\gamma_Z\rightarrow t_L\bar t_Lg_R)&=&
\phantom{-}{A_{+-}\over\sqrt{2}}
\cos{\theta_{tg}\over2}
\bigl(x_t\beta_t\sin{\theta_{tg}\over2}\sin{\theta_{t\bar t}\over2}
+(1-\bar x_t)\cos{\theta_{t\bar t}+\theta_{tg}\over2}\bigr)\nonumber\\
{\cal M}(L;\gamma_L\rightarrow t_R\bar t_Lg_L)&=&
-A_{--}
\cos^2{\theta_{tg}\over2}\cos{\theta_{t\bar t}\over2}
(1-x_t)\nonumber\\
{\cal M}(L;\gamma_R\rightarrow t_R\bar t_Lg_L)&=&
-A_{--}
\sin{\theta_{tg}\over2}\cos{\theta_{tg}\over2}
\sin{\theta_{t\bar t}\over2}\bigl(x_t\beta_t-(1-x_t)\bigr)
\nonumber\\
{\cal M}(L;\gamma_Z\rightarrow t_R\bar t_Lg_L)&=&
-{A_{--}\over\sqrt{2}}
\cos{\theta_{tg}\over2}
\bigl(x_t\beta_t\sin{\theta_{tg}\over2}\cos{\theta_{t\bar t}\over2}
+(1-x_t)\sin{\theta_{t\bar t}-\theta_{tg}\over2}\bigr)\nonumber\\
{\cal M}(L;\gamma_L\rightarrow t_R\bar t_Lg_R)&=&
\phantom{-}0\nonumber\\
{\cal M}(L;\gamma_R\rightarrow t_R\bar t_Lg_R)&=&
\phantom{-}A_{--}\cos{\theta_{tg}\over2}
\bigl(x_t\beta_t\sin{\theta_{tg}\over2}
\sin{\theta_{t\bar t}\over2}+(1-\bar x_t)
\cos{\theta_{t\bar t}+\theta_{tg}\over2}\bigr)
\nonumber\\
{\cal M}(L;\gamma_Z\rightarrow t_R\bar t_Lg_R)&=&
\phantom{-}{A_{--}\over\sqrt{2}}
\sin{\theta_{tg}\over2}
\bigl(x_t\beta_t\cos{\theta_{tg}\over2}\cos{\theta_{t\bar t}\over2}
+(1-\bar x_t)\cos{\theta_{t\bar t}+\theta_{tg}\over2}\bigr)\nonumber\\
{\cal M}(L;\gamma_L\rightarrow t_L\bar t_Rg_L)&=&
-A_{++}
\sin{\theta_{tg}\over2}\cos{\theta_{tg}\over2}
\sin{\theta_{t\bar t}\over2}\bigl(x_t\beta_t+(1-x_t)\bigr)
                                                        \label{ttgamps} \\
{\cal M}(L;\gamma_R\rightarrow t_L\bar t_Rg_L)&=&
-A_{++}
\sin^2{\theta_{tg}\over2}\cos{\theta_{t\bar t}\over2}
(1-x_t)\nonumber\\
{\cal M}(L;\gamma_Z\rightarrow t_L\bar t_Rg_L)&=&
\phantom{-}{A_{++}\over\sqrt{2}}
\sin{\theta_{tg}\over2}
\bigl(x_t\beta_t\cos{\theta_{tg}\over2}\cos{\theta_{t\bar t}\over2}
+(1-x_t)\cos{\theta_{t\bar t}-\theta_{tg}\over2}\bigr)\nonumber\\
{\cal M}(L;\gamma_L\rightarrow t_L\bar t_Rg_R)&=&
\phantom{-}A_{++}\sin{\theta_{tg}\over2}
\bigl(x_t\beta_t\cos{\theta_{tg}\over2}
\sin{\theta_{t\bar t}\over2}+(1-\bar x_t)
\sin{\theta_{t\bar t}+\theta_{tg}\over2}\bigr)
\nonumber\\
{\cal M}(L;\gamma_R\rightarrow t_L\bar t_Rg_R)&=&
\phantom{-}0\nonumber\\
{\cal M}(L;\gamma_Z\rightarrow t_L\bar t_Rg_R)&=&
-{A_{++}\over\sqrt{2}}
\cos{\theta_{tg}\over2}
\bigl(x_t\beta_t\sin{\theta_{tg}\over2}\cos{\theta_{t\bar t}\over2}
-(1-\bar x_t)\sin{\theta_{t\bar t}+\theta_{tg}\over2}\bigr)\nonumber\\
{\cal M}(L;\gamma_L\rightarrow t_R\bar t_Rg_L)&=&
-A_{-+}
\cos^2{\theta_{tg}\over2}\sin{\theta_{t\bar t}\over2}(1-x_t)
\nonumber\\
{\cal M}(L;\gamma_R\rightarrow t_R\bar t_Rg_L)&=&
\phantom{-}A_{-+}
\sin{\theta_{tg}\over2}\cos{\theta_{tg}\over2}\cos{\theta_{t\bar t}\over2}
\bigl(x_t\beta_t-(1-x_t)\bigr)\nonumber\\
{\cal M}(L;\gamma_Z\rightarrow t_R\bar t_Rg_L)&=&
-{A_{-+}\over\sqrt{2}}
\cos{\theta_{tg}\over2}
\bigl(x_t\beta_t\sin{\theta_{tg}\over2}\sin{\theta_{t\bar t}\over2}
-(1-x_t)\cos{\theta_{t\bar t}-\theta_{tg}\over2}\bigr)\nonumber\\
{\cal M}(L;\gamma_L\rightarrow t_R\bar t_Rg_R)&=&
\phantom{-}0\nonumber\\
{\cal M}(L;\gamma_R\rightarrow t_R\bar t_Rg_R)&=&
-A_{-+}\cos{\theta_{tg}\over2}
\bigl(x_t\beta_t\sin{\theta_{tg}\over2}\cos{\theta_{t\bar t}\over2}
-(1-\bar x_t)\sin{\theta_{t\bar t}+\theta_{tg}\over2}\bigr)\nonumber\\
{\cal M}(L;\gamma_Z\rightarrow t_R\bar t_Rg_R)&=&
\phantom{-}{A_{-+}\over\sqrt{2}}
\sin{\theta_{tg}\over2}
\bigl(x_t\beta_t\cos{\theta_{tg}\over2}\sin{\theta_{t\bar t}\over2}
+(1-\bar x_t)\sin{\theta_{t\bar t}+\theta_{tg}\over2}\bigr)\ ,
\nonumber
\end{eqnarray}
where
\begin{equation}
A_{\pm\pm} \ =\ -ie^2g_sT^a\,{x_g\Bigl[x_t\bar x_t(1\pm\beta_t)
(1\pm\bar\beta_t)\Bigr]^{1/2}\over\sqrt{s}(1-x_t)(1-\bar x_t)}\ ,
\label{Apmpm}
\end{equation}
with ${\rm Tr}(T^aT^b) =\delta^{ab}/2$.
The remaining matrix elements can be obtained from
\begin{eqnarray}
{\cal M}(L,R;\gamma_\lambda\rightarrow t_L\bar t_L g_\sigma)&=&
 -(-1)^\lambda{\cal M}(R,L;\gamma_{-\lambda}\rightarrow t_R\bar t_R
g_{-\sigma})\nonumber\\
{\cal M}(L,R;\gamma_\lambda\rightarrow t_L\bar t_R g_\sigma)&=&
\phantom{-}(-1)^\lambda{\cal M}(R,L;\gamma_{-\lambda}\rightarrow
t_R\bar t_L g_{-\sigma})\ .\label{Ramps}
\end{eqnarray}
In terms of the variables in equation (\ref{prodcut}) the energy
fractions are
\begin{eqnarray}
x_g&=& x\beta^2\nonumber\\
x_t&=& 1-{x_g\over2}+x_g(\Delta-{1\over2})
\biggl({\beta^2-x_g
\over1-x_g}\biggr)^{1/2}\label{xs}\\
\bar x_t&=& 2-x_g-x_t\nonumber
\end{eqnarray}
Here $\beta^2=1-4m_t^2/s$ is the tree level velocity of the top
quarks, while the
velocities of the $t$ and $\bar t$ in the presence of the radiated
gluon are
\begin{eqnarray}
\beta_t^2 &=&1 -{4m_t^2\over x_t^2s}\nonumber\\
\bar\beta_t^2&=&1 -{4m_t^2\over \bar x_t^2s}
\ .\label{velocity}
\end{eqnarray}
The lab-frame angles are obtained from
\begin{eqnarray}
 \cos\theta_{t\bar t}&=& {1\over x_t\beta_t\bar x_t\bar\beta_t}\bigl[
 x_g-x_t-\bar x_t+x_t\bar x_t+{4m_t^2\over s}\bigr]\nonumber\\
 \cos\theta_{tg}&=&{1\over x_t\beta_tx_g}\bigl[\bar x_t-x_t-x_g
 +x_tx_g\bigr]\ .\label{tangle}
\end{eqnarray}

\section{$t\rightarrow bW^+g$ decay amplitudes}

The helicity amplitudes for top decay with a radiated gluon can be
calculated in an analogous manner to the production calculation in
appendix A.  We describe the decay configuration in the top
rest frame in terms of five variables.  Two of these are the energy
fractions $x_i=2E_i/m_t$ of the $W^+$ and the gluon.  These energies
are determined by the variables of equation (\ref{decaycut}), and
they fix all the energies and angles within the $bW^+g$ decay plane.
Two more variables are the polar angle $\chi_t$ and azimuthal angle
$\psi_t$ of the $W^+$ with respect to the top momentum boost axis.
The final variable is the angle $\phi_g$ between the plane given by
the top boost axis and the $W^+$ momentum and the $bW^+g$-plane,
rotated around the $W^+$ momentum.  This rotation by $\phi_g$ also
rotates the $W^+$ decay products.

We can make explicit the dependence of the
matrix elements on the variables $\chi_t$, $\psi_t$, and $\phi_g$ if we
expand the top helicity eigenstates $t_h$ onto a basis of spin
eigenstates along the $W^+$ momentum direction.  Labeling these states
as $t^\prime_h$ we obtain the relations:
\begin{eqnarray}
\lefteqn{{\cal M}(t_L\rightarrow bW^+g)
\ =\ e^{-i\psi_t/2}\Bigl[}\hspace{.5in} \nonumber\\
&&{\cal M}(t^\prime_R\rightarrow bW^+g)\sin{\chi_t\over2}e^{i\phi_g/2}
+{\cal M}(t^\prime_L\rightarrow bW^+g)\cos{\chi_t\over2}e^{-i\phi_g/2}
\Bigr]\nonumber\\
\lefteqn{{\cal M}(t_R\rightarrow bW^+g)
\ =\ e^{i\psi_t/2}\Bigl[}\hspace{.5in}                   \label{bWg}\\
&&{\cal M}(t^\prime_R\rightarrow bW^+g)\cos{\chi_t\over2}e^{i\phi_g/2}
-{\cal M}(t^\prime_L\rightarrow bW^+g)\sin{\chi_t\over2}e^{-i\phi_g/2}
\Bigr]\ .\nonumber
\end{eqnarray}
The helicity amplitudes in this basis are
\begin{eqnarray}
{\cal M}(t^\prime_R\rightarrow b_LW_Rg_L)&=&
-{2\over\sqrt{\zeta x_g}}\Bigl(x_g\cos{\theta_{Wg}\over2}
+x_b\cos{\theta_{Wb}\over2}\cos{\theta_{Wg}+\theta_{Wb}\over2}
\Bigr)\nonumber\\
{\cal M}(t^\prime_L\rightarrow b_LW_Rg_L)&=&
\phantom{-}0\nonumber\\
{\cal M}(t^\prime_R\rightarrow b_LW_Lg_L)&=&
\phantom{-}0\nonumber\\
{\cal M}(t^\prime_L\rightarrow b_LW_Lg_L)&=&\phantom{-}
{2\over\sqrt{\zeta x_g}}\Bigl(-x_g\sin{\theta_{Wg}\over2}
+x_b\sin{\theta_{Wb}\over2}\cos{\theta_{Wg}+\theta_{Wb}\over2}
\Bigr)\nonumber\\
{\cal M}(t^\prime_R\rightarrow b_LW_Zg_L)&=&\phantom{-}
{x_W(1+\beta_W)\over w\sqrt{2\zeta x_g}}\Bigl(-x_g\sin{\theta_{Wg}\over2}
+x_b\sin{\theta_{Wb}\over2}\cos{\theta_{Wg}+\theta_{Wb}\over2}
\Bigr)\nonumber\\
{\cal M}(t^\prime_L\rightarrow b_LW_Zg_L)&=&
-{x_W(1-\beta_W)\over w\sqrt{2\zeta x_g}}\Bigl(x_g\cos{\theta_{Wg}\over2}
+x_b\cos{\theta_{Wb}\over2}\cos{\theta_{Wg}+\theta_{Wb}\over2}
\Bigr)\nonumber\\
{\cal M}(t^\prime_R\rightarrow b_LW_Rg_R)&=&\phantom{-}
\sqrt{x_b}\cos{\theta_{Wb}\over2}\Bigl(2\sqrt{x_b\over\zeta x_g}
\cos{\theta_{Wg}+\theta_{Wb}\over2} - \sin{\theta_{Wg}}
\Bigr)\nonumber\\
{\cal M}(t^\prime_L\rightarrow b_LW_Rg_R)&=&
-\sqrt{x_b}\cos{\theta_{Wb}\over2}(1-\cos{\theta_{Wg}})
                                                       \label{WBG}\\
{\cal M}(t^\prime_R\rightarrow b_LW_Lg_R)&=&
-\sqrt{x_b}\sin{\theta_{Wb}\over2}(1+\cos{\theta_{Wg}})
\nonumber\\
{\cal M}(t^\prime_L\rightarrow b_LW_Lg_R)&=&
-\sqrt{x_b}\sin{\theta_{Wb}\over2}\Bigl(2\sqrt{x_b\over\zeta x_g}
\cos{\theta_{Wg}+\theta_{Wb}\over2} + \sin{\theta_{Wg}}
\Bigr)\nonumber\\
{\cal M}(t^\prime_R\rightarrow b_LW_Zg_R)&=&
{x_W\sqrt{x_b}\over2\sqrt{2}w}\biggl[
(1+\beta_W)\sin{\theta_{Wb}\over2}\Bigl(-2\sqrt{x_b\over\zeta x_g}
\cos{\theta_{Wg}+\theta_{Wb}\over2} + \sin{\theta_{Wg}}\Bigr)
\nonumber\\
&&\qquad\quad
+(1-\beta_W)\cos{\theta_{Wb}\over2}(1+\cos{\theta_{Wg}})
\biggr]\nonumber\\
{\cal M}(t^\prime_L\rightarrow b_LW_Zg_R)&=&
{x_W\sqrt{x_b}\over2\sqrt{2}w}\biggl[
(1-\beta_W)\cos{\theta_{Wb}\over2}\Bigl(2\sqrt{x_b\over\zeta x_g}
\cos{\theta_{Wg}+\theta_{Wb}\over2} + \sin{\theta_{Wg}}\Bigr)
\nonumber\\
&&\qquad\quad
+(1+\beta_W)\sin{\theta_{Wb}\over2}(1-\cos{\theta_{Wg}})
\biggr]\ .\nonumber
\end{eqnarray}
where we have dropped a factor of $-iT^ag_sg/\sqrt{2}$.
In terms of the variables of (\ref{decaycut}) the energy fractions are
\begin{eqnarray}
x_g&=& y(1-w^2)\nonumber\\
x_W&=& 1+w^2\,-\,z\,{x_g(1-w^2-x_g)\over1-x_g}              \label{ys}\\
x_b&=& 2-x_g-x_W\ ,\nonumber
\end{eqnarray}
and we have also introduced the variable $\zeta=2p_b\cdot p_g/m_t^2
=1+w^2-x_W$.  The velocity of the $W^+$ is given by
\begin{equation}
\beta_W^2\ =\ 1 -{4w^2\over x_W^2}  \ .             \label{Wvelocity}
\end{equation}
The angles in the top rest frame are obtained from
\begin{eqnarray}
 \cos\theta_{Wb}&=& {1\over x_W\beta_Wx_b}\bigl[
 x_g-x_W-x_b+x_Wx_b+2w^2\bigr]\nonumber\\
 \cos\theta_{Wg}&=& {1\over x_W\beta_Wx_g}\bigl[
 x_b-x_W-x_g+x_Wx_g+2w^2\bigr]\ .\label{Wangle}
\end{eqnarray}

The amplitudes for $\bar t$ decay in its rest frame
can be obtained from these by simply using
\begin{equation}
{\cal M}(t_h\rightarrow b_{\rho}W^+_\lambda g_\sigma)\ =\
{\cal M}(\bar t_{-h}\rightarrow \bar b_{-\rho}W^-_{-\lambda} g_-\sigma)\ ,
\label{antitg}
\end{equation}
while replacing all of the energies and polar angles of $t$ decay with
the corresponding variables of $\bar t$ decay and replacing the
azimuthal angles by $\psi_t\rightarrow-\bar\psi_t$ and
$\phi_g\rightarrow-\phi_g$.

\newpage

\noindent
\vskip8cm
\begin{displaymath}
\vbox{\tabskip=0pt \offinterlineskip
\def\tablerule{\noalign{\hrule}}
\halign{\vphantom{\Big[}\strut#&\vrule#&\hfil #\hfil &&\vrule#&
\quad \hfil #\quad \cr
\tablerule
&&\multispan2
&\multispan7 \hfil $\mu_{cut}$ (in GeV) \hfil && {\rm production}\hfil &\cr
&&\multispan2
&\multispan2 \hfil 5\hfil
&\multispan2 \hfil 10\hfil
&\multispan2 \hfil 20\hfil & $\infty$\hfil &&
 {\rm only}\hfil &\cr\tablerule
&&$\ e^-_L\ $
&& -2.8
&& -0.7
&& +1.2
&& +3.3
&& +3.2 \hfil&\cr\tablerule
&&$\ e^-_R\ $
&& +4.2
&& +3.6
&& +3.0
&& +2.5
&& +2.9 \hfil&\cr\tablerule}}
\end{displaymath}
\baselineskip=12pt
Table 1: Percentage ${\cal O}(\alpha_s)$ corrections to the
top quark forward-backward asymmetry for $m_t=175$ GeV and
$\sqrt{s}=400$ GeV with polarized electrons.  The first four columns
are using the reconstruction algorithm (\ref{mu}), while the last
column gives the corrections from production only, assuming exact
event reconstruction.

\newpage

\begin{figure}
\vskip-4cm
\epsfysize=16cm
\centerline{\epsffile{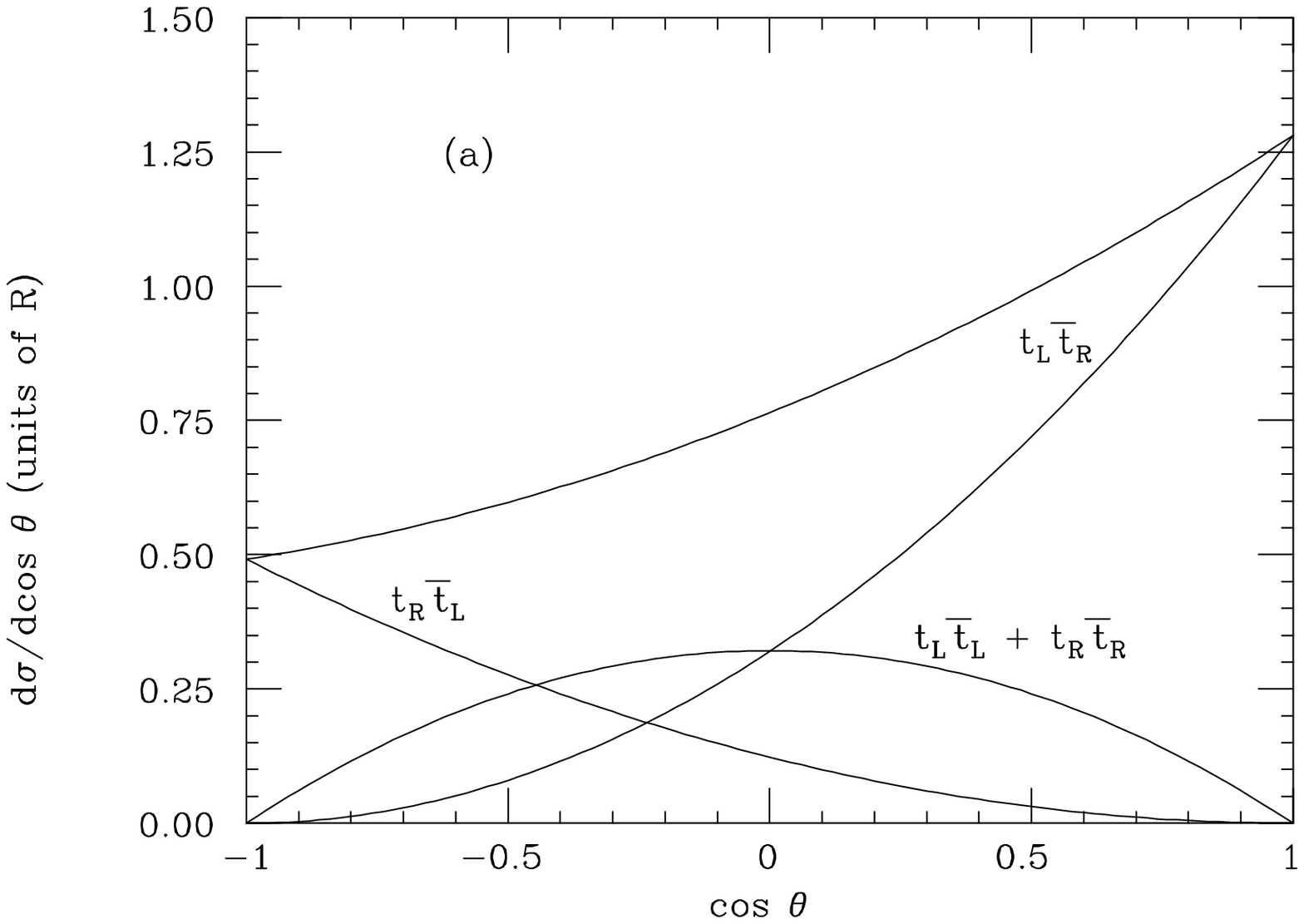}}
\vskip-4cm
\vskip-4cm
\epsfysize=16cm
\centerline{\epsffile{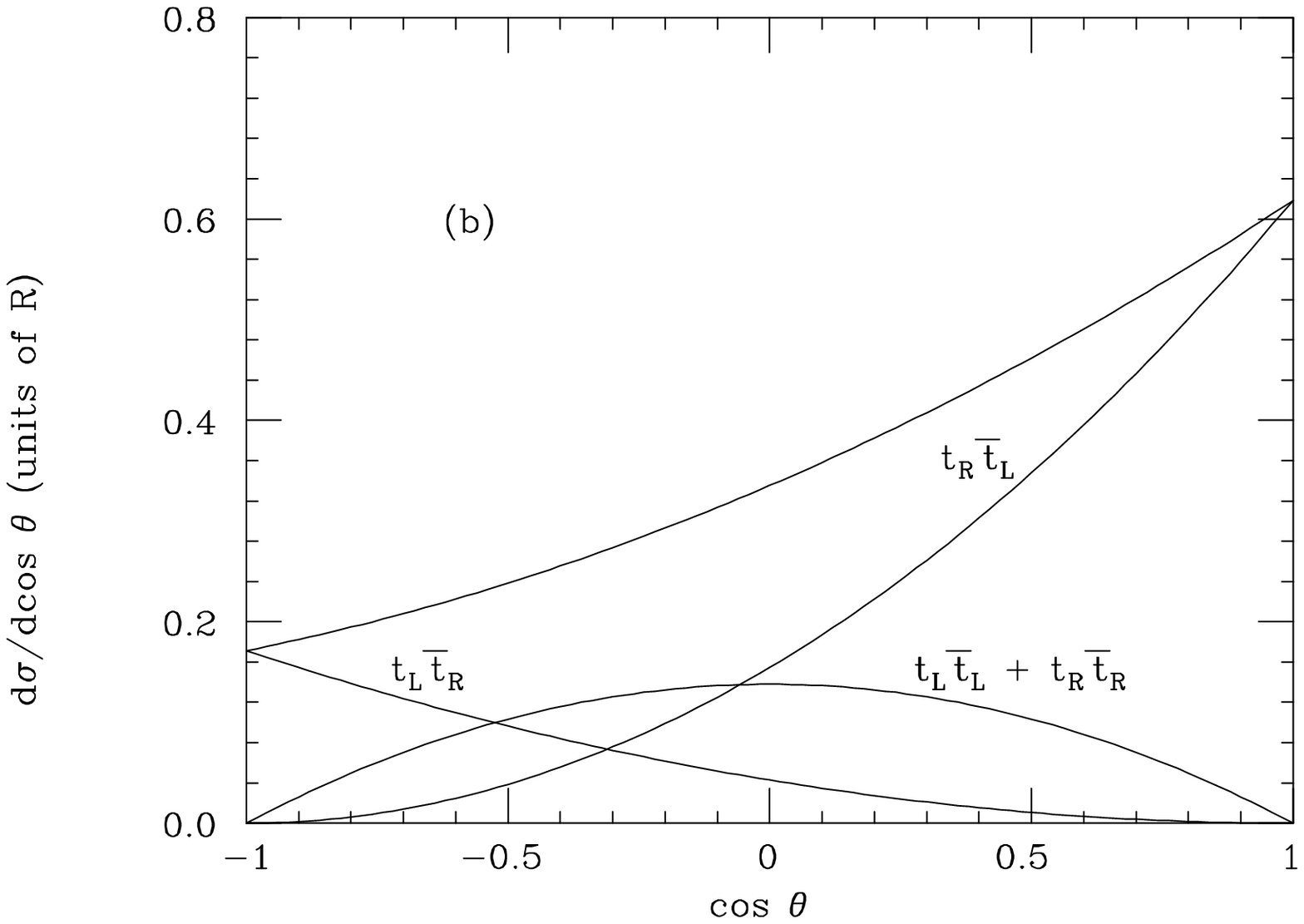}}
\vskip-4cm
\vskip6pt
\baselineskip=12pt
Fig.~1: $e^+e^-\rightarrow t\bar t$ cross section for
(a) left-polarized electrons and
(b) right-polarized electrons.
\end{figure}

\newpage

\begin{figure}
\vskip-6cm
\epsfysize=16cm
\centerline{\epsffile{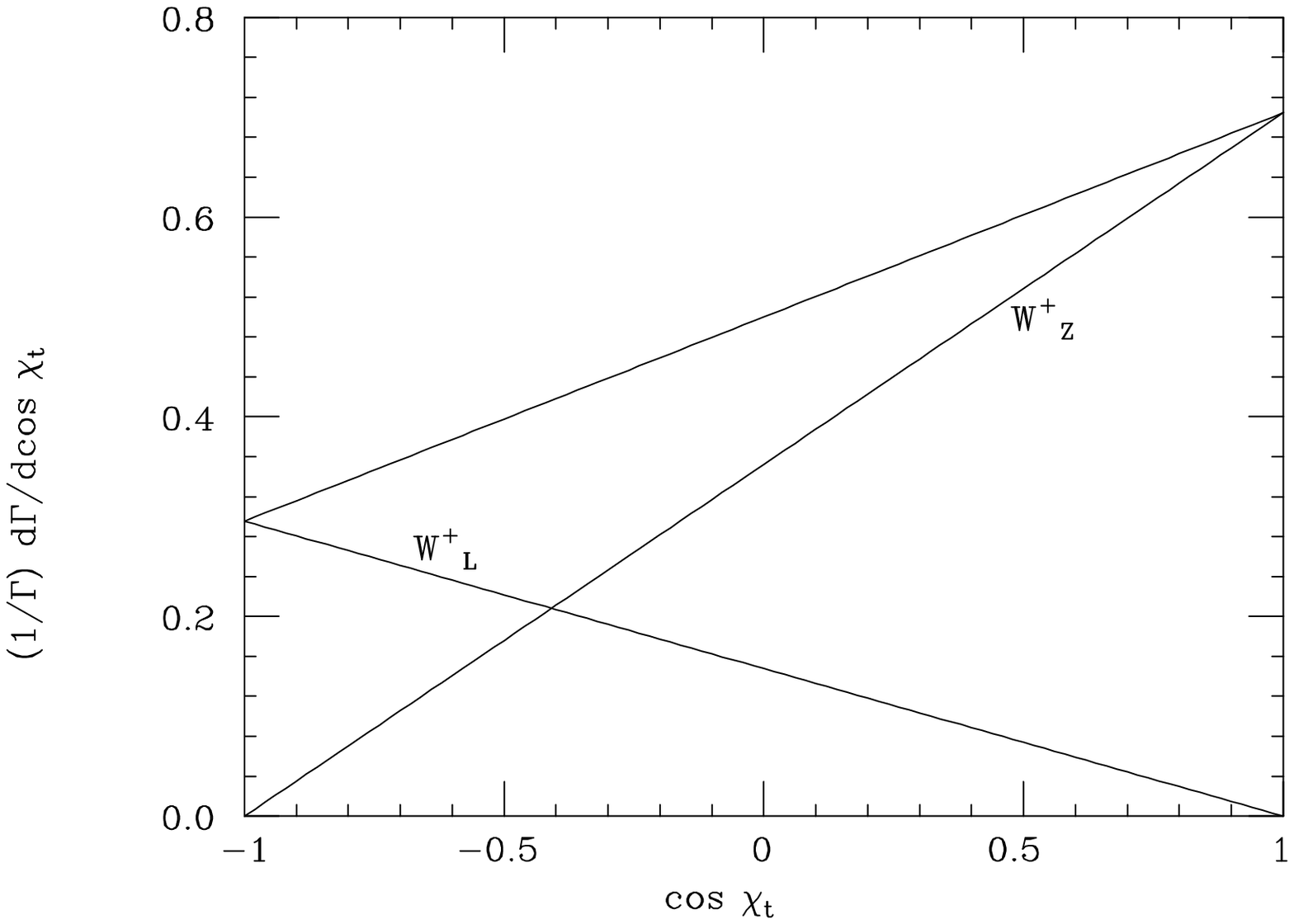}}
\vskip-4cm
\vskip6pt
\baselineskip=12pt
Fig.~2: Polar angle dependence of $W^+$ from decay of right-handed top
quark.
\vskip-3cm
\epsfysize=16cm
\centerline{\epsffile{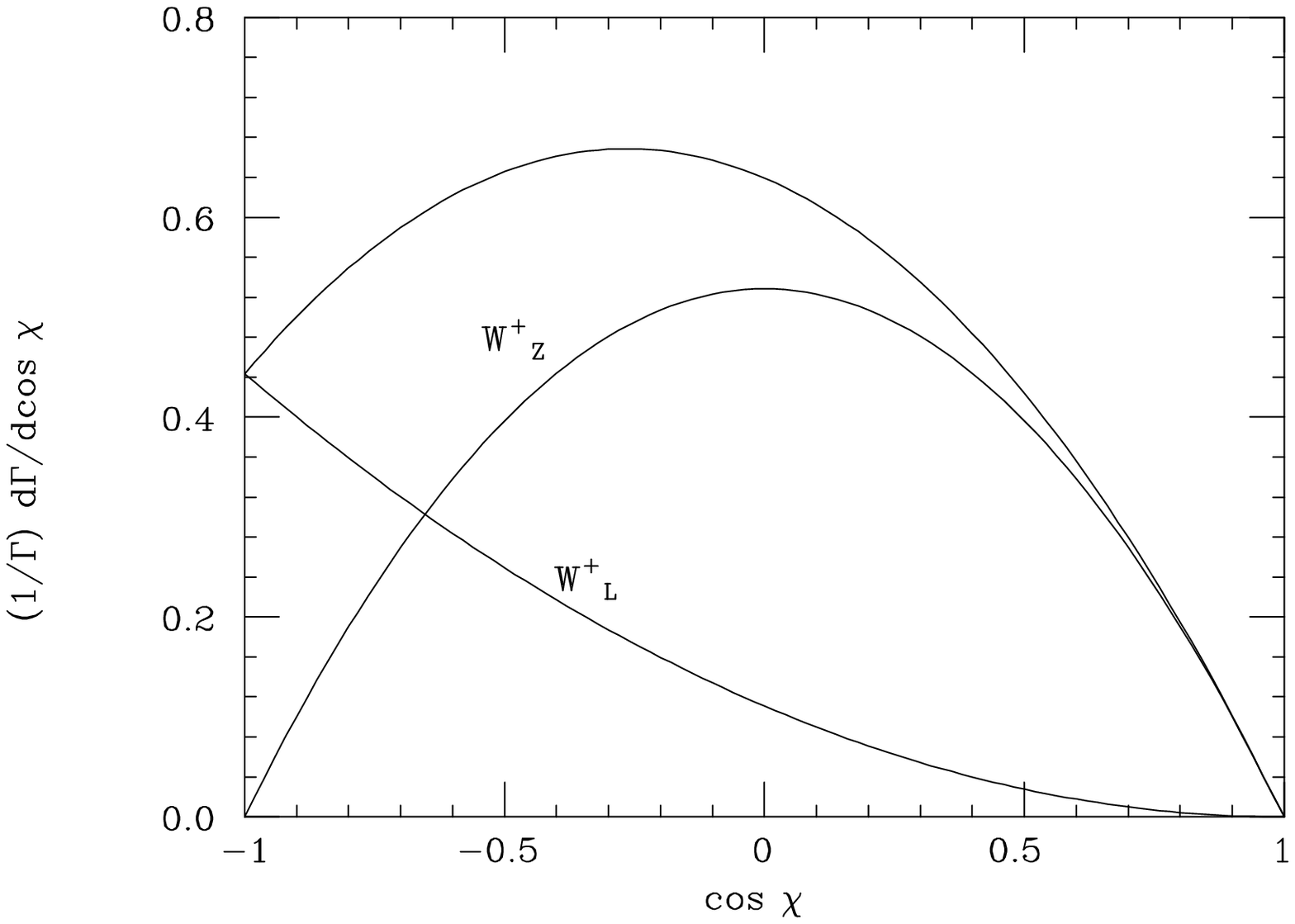}}
\vskip-4cm
\vskip6pt
\baselineskip=12pt
Fig~3: Polar angle dependence of $\ell^+$ from decay of $W^+$ in a
$t\bar t$ event.
\end{figure}

\newpage

\begin{figure}
\vskip-4cm
\epsfysize=16cm
\centerline{\epsffile{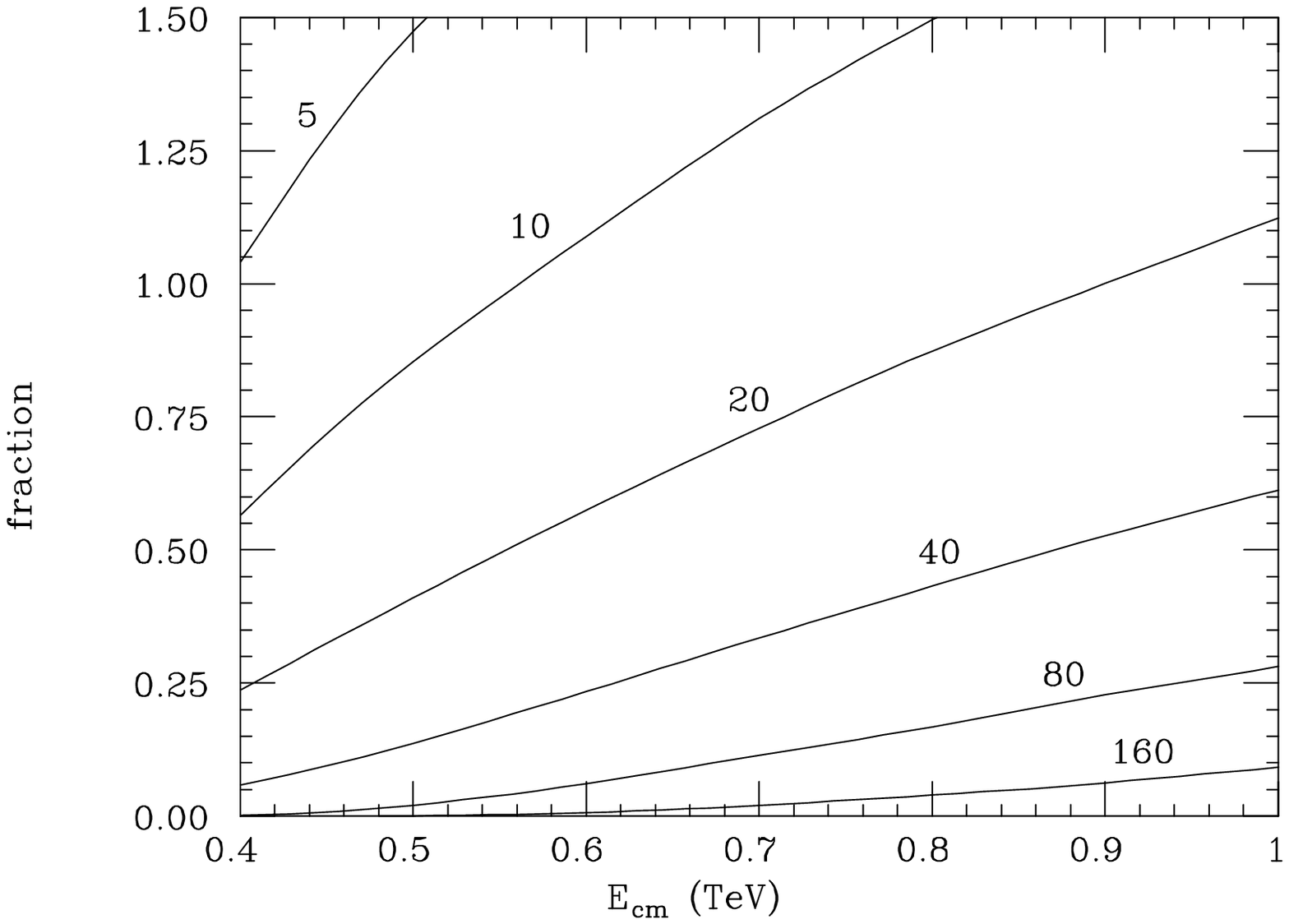}}
\vskip-4cm
\vskip6pt
\baselineskip=12pt
Fig.~4: Fraction of events containing a production gluon as a function
of $\sqrt{s}$.  The curves are, from top to bottom, for $\mu_{cut}=5$,
10, 20, 40, 80, and 160 GeV.
\end{figure}

\newpage

\begin{figure}
\vskip-4cm
\epsfysize=16cm
\centerline{\epsffile{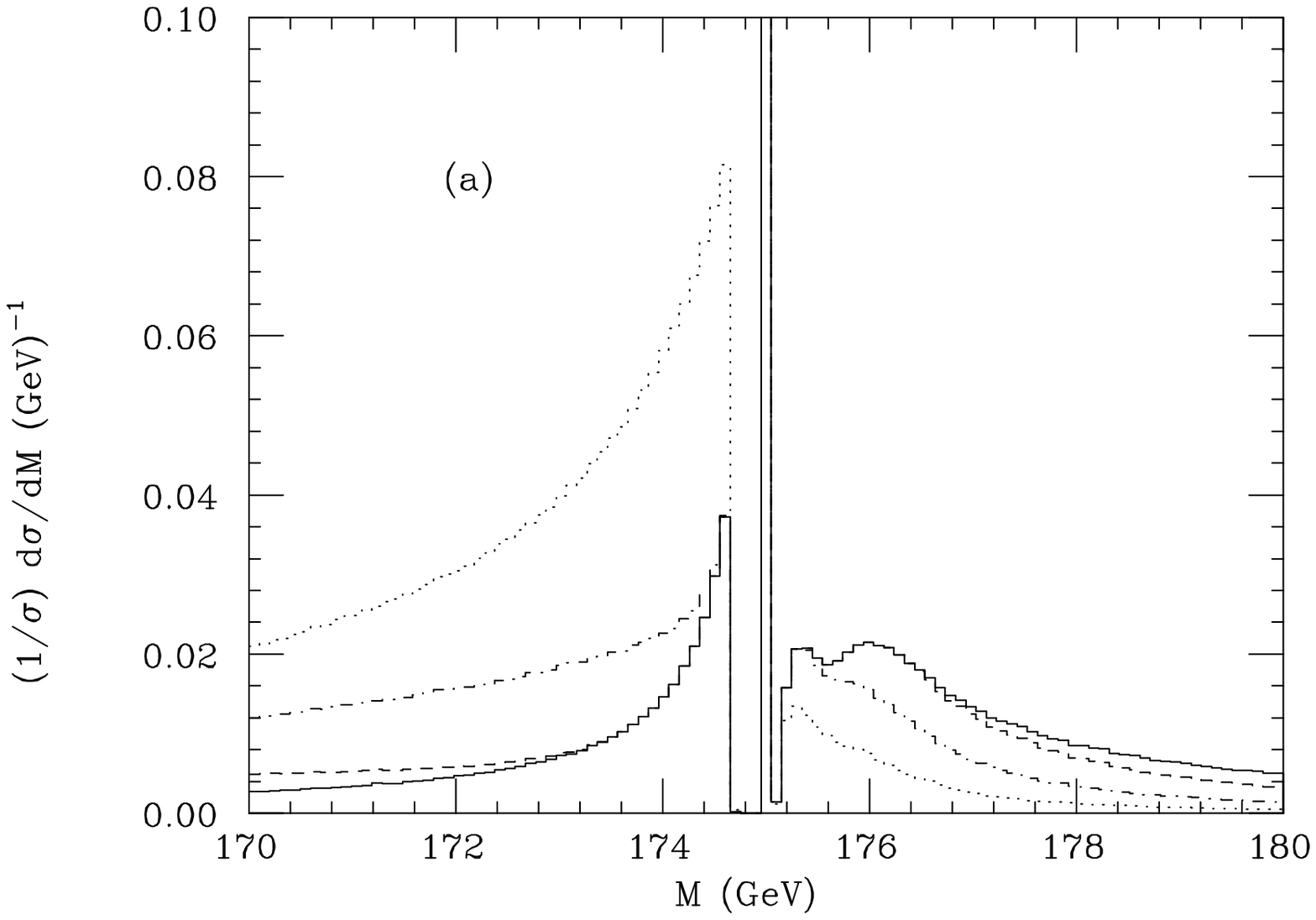}}
\vskip-4cm
\vskip-4cm
\epsfysize=16cm
\centerline{\epsffile{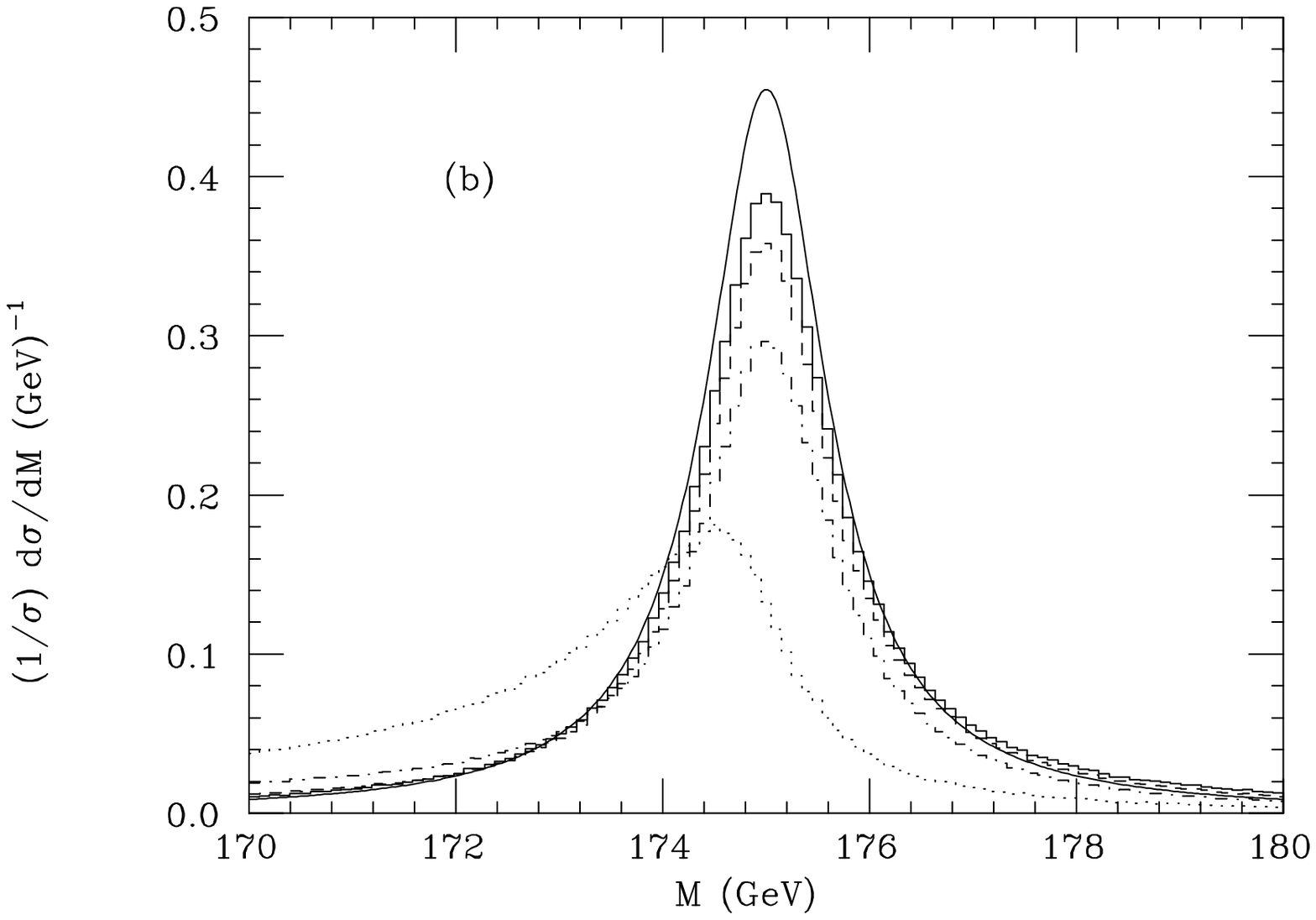}}
\vskip-4cm
\vskip6pt
\baselineskip=12pt
Fig.~5: Top mass reconstruction distributions for $\sqrt{s}=400$ GeV
(a) in the zero-width limit and
(b) with an initial Breit-Wigner resonance distribution.
The histograms are for $\mu_{cut}=5$ GeV (dots), 10 GeV (dotdash),
20 GeV (dashes), and $\infty$ (solid).  The smooth curve in (b)
is the original Breit-Wigner distribution.
\end{figure}

\newpage

\begin{figure}
\vskip-4cm
\epsfysize=16cm
\centerline{\epsffile{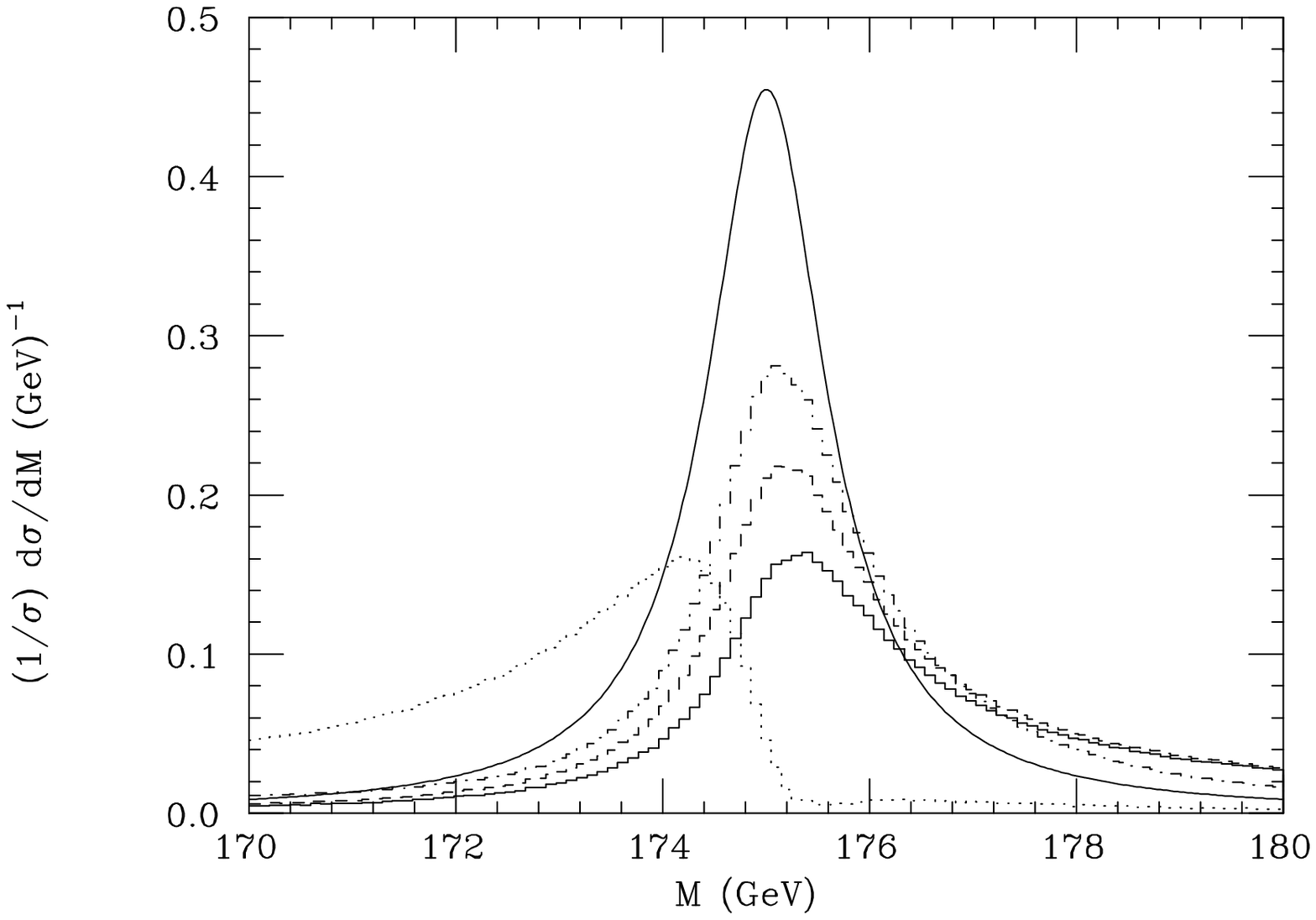}}
\vskip-4cm
\vskip6pt
\baselineskip=12pt
Fig.~6: Top mass reconstruction distributions for $\sqrt{s}=1$ TeV
with an initial Breit-Wigner resonance distribution.
The histograms are for $\mu_{cut}=5$ GeV (dots), 20 GeV (dotdash),
80 GeV (dashes), and $\infty$ (solid).  The smooth curve
is the original Breit-Wigner distribution.
\end{figure}

\newpage

\begin{figure}
\vskip-4cm
\epsfysize=16cm
\centerline{\epsffile{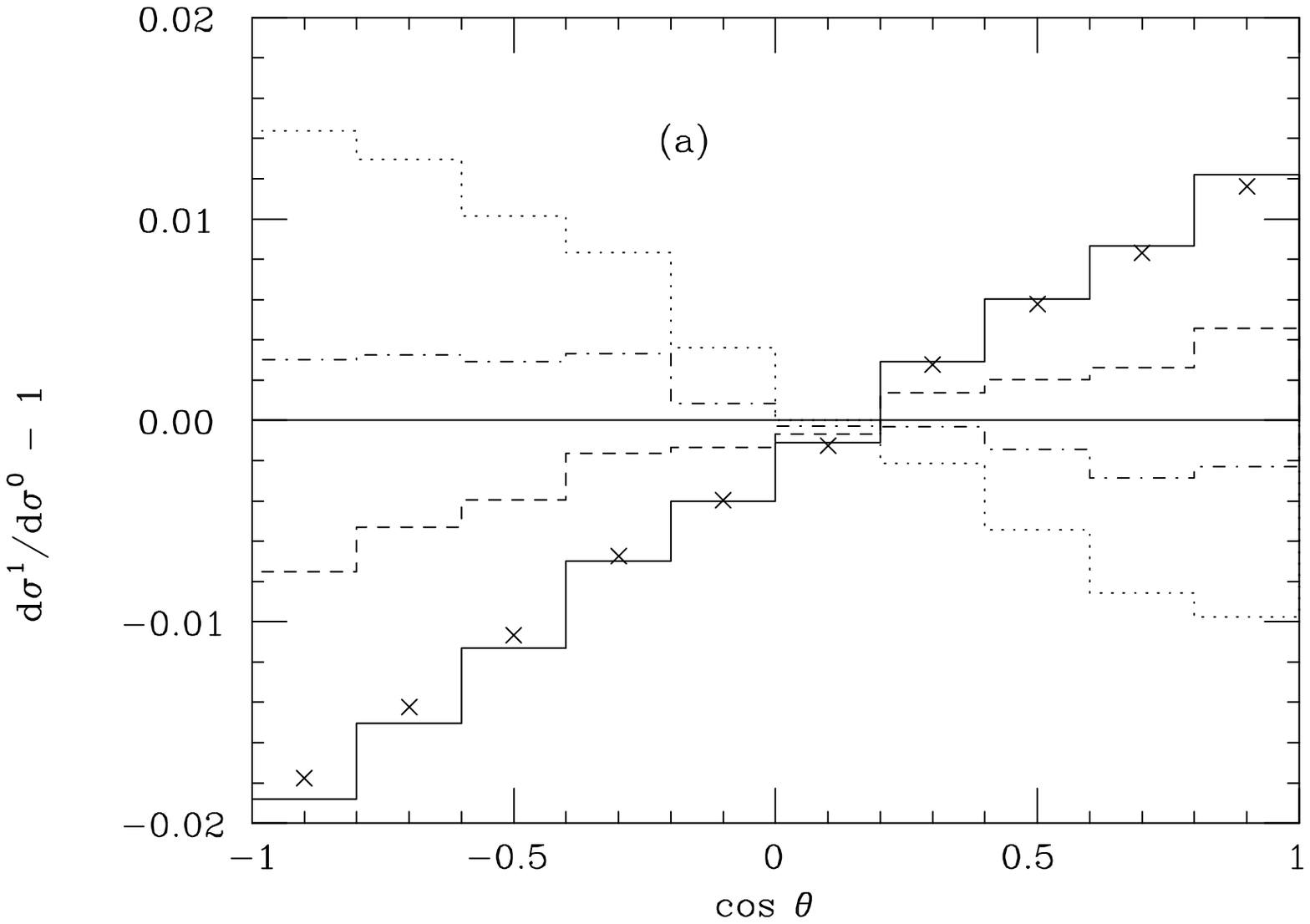}}
\vskip-4cm
\vskip-4cm
\epsfysize=16cm
\centerline{\epsffile{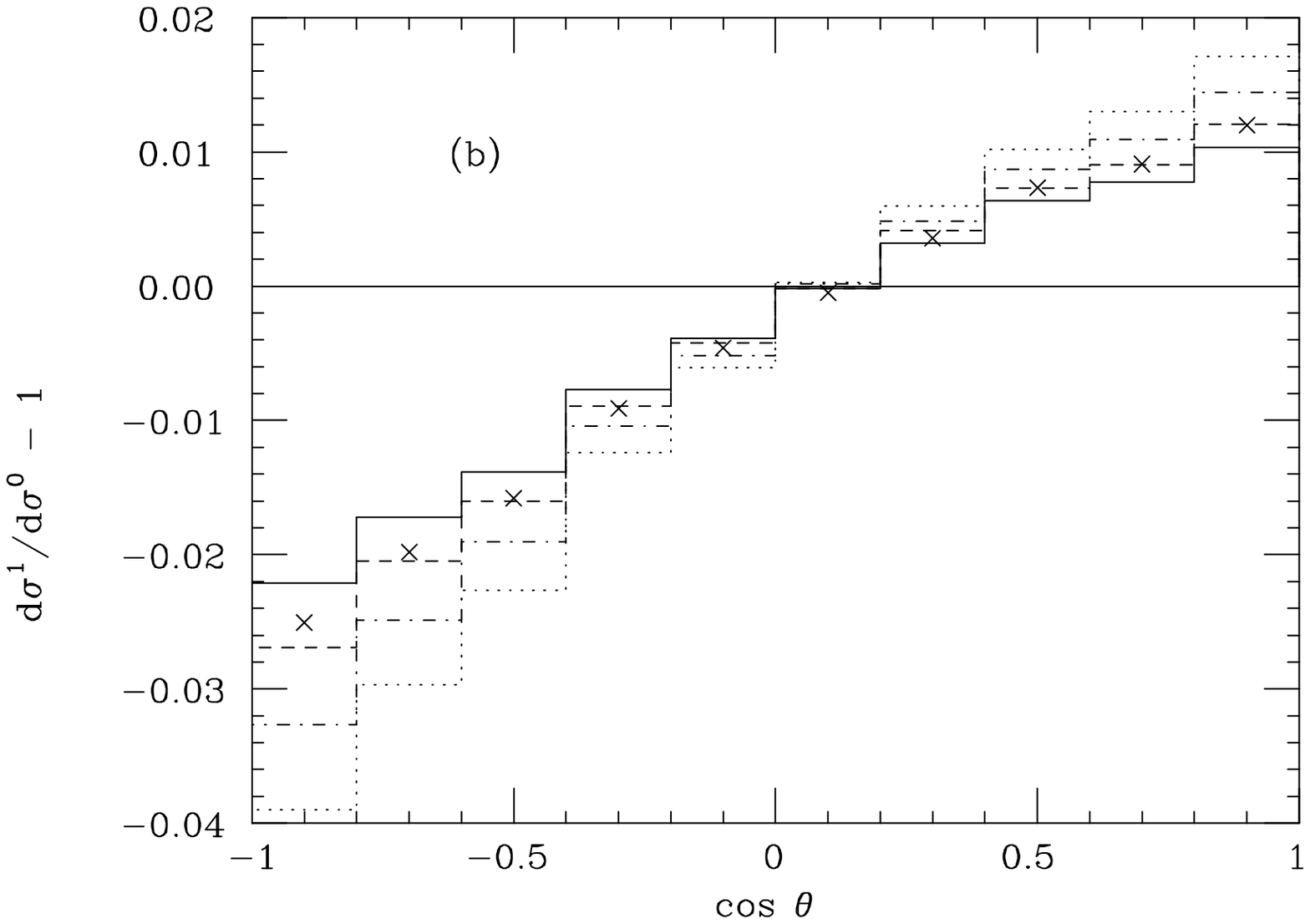}}
\vskip-4cm
\vskip6pt
\baselineskip=12pt
Fig.~7: The ${\cal O}(\alpha_s)$ corrections to the top quark polar angle
distributions for $\sqrt{s}=400$ GeV with
(a) left-polarized electrons and
(b) right-polarized electrons.
The histograms are for $\mu_{cut}=5$ GeV (dots), 10 GeV (dotdash),
20 GeV (dashes), and $\infty$ (solid), while the points plotted with
the symbol $\times$ are the pure production corrections.

\end{figure}

\newpage

\begin{figure}
\vskip-4cm
\epsfysize=16cm
\centerline{\epsffile{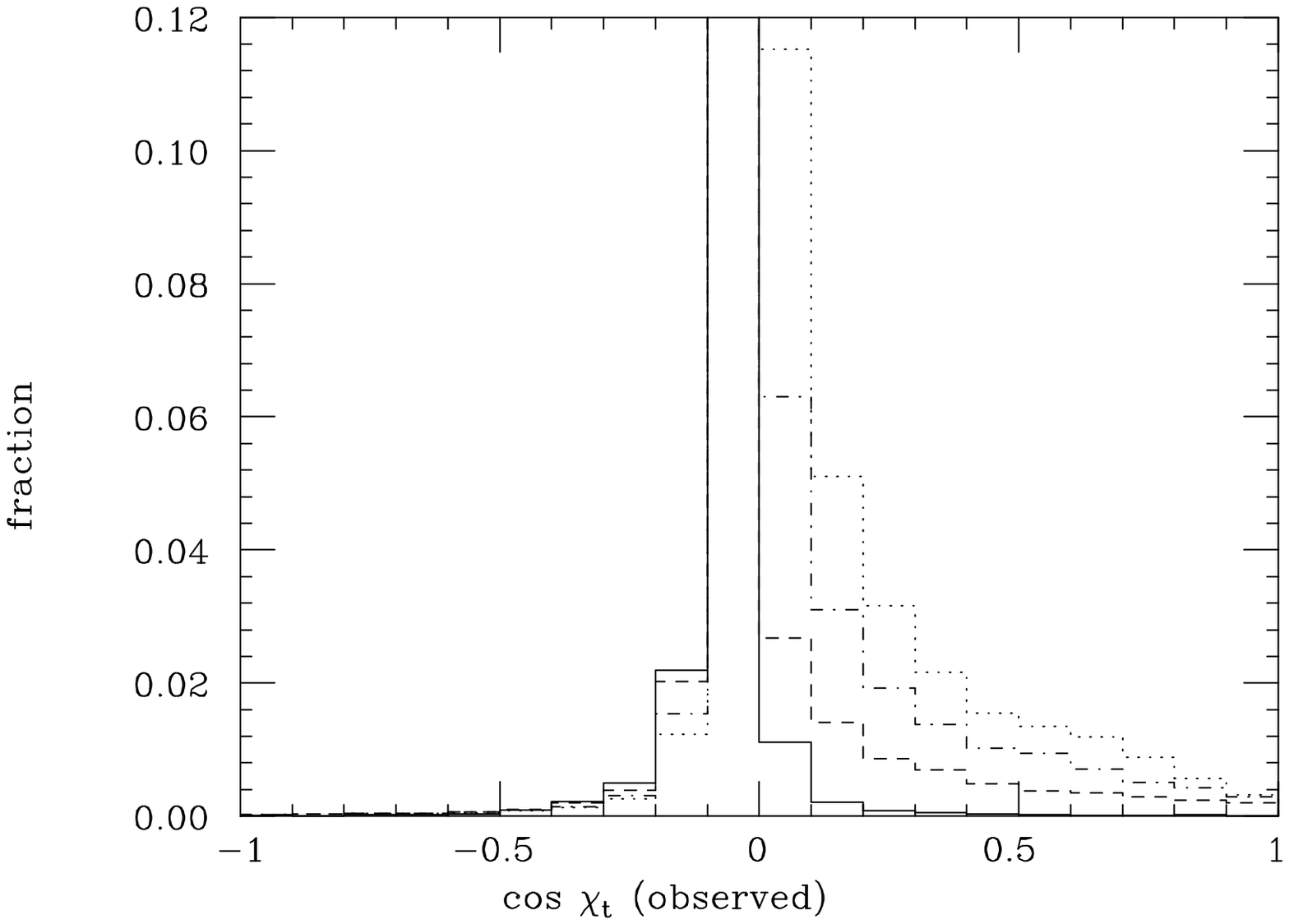}}
\vskip-4cm
\vskip6pt
\baselineskip=12pt
Fig.~8: Distribution of observed $\cos\chi_t$ for events with true
$\cos\chi_t$ between -0.1 and 0.0.
The histograms are for $\mu_{cut}=5$ GeV (dots), 10 GeV (dotdash),
20 GeV (dashes), and $\infty$ (solid).
\end{figure}

\newpage

\begin{figure}
\vskip-4cm
\epsfysize=16cm
\centerline{\epsffile{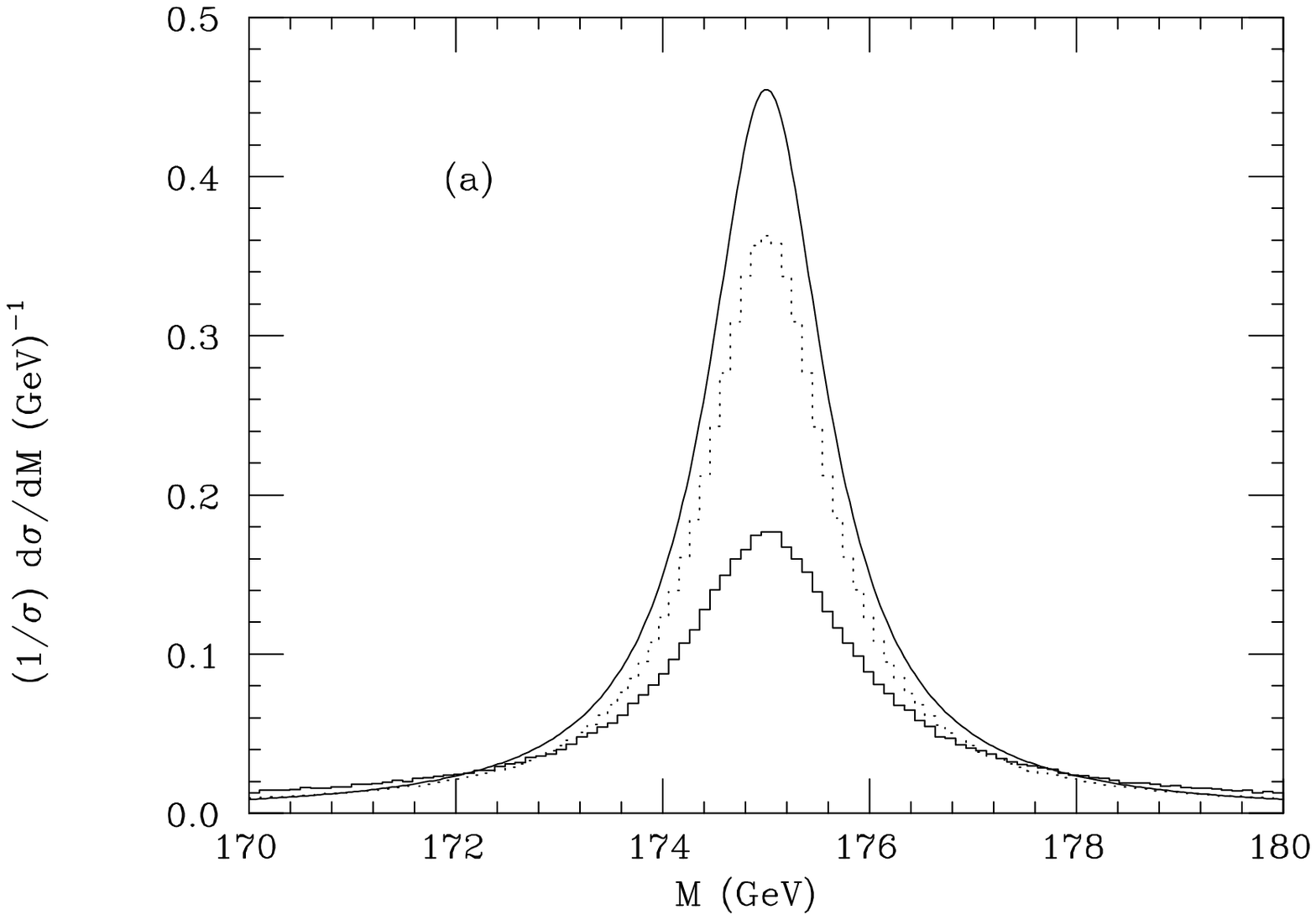}}
\vskip-4cm
\vskip-4cm
\epsfysize=16cm
\centerline{\epsffile{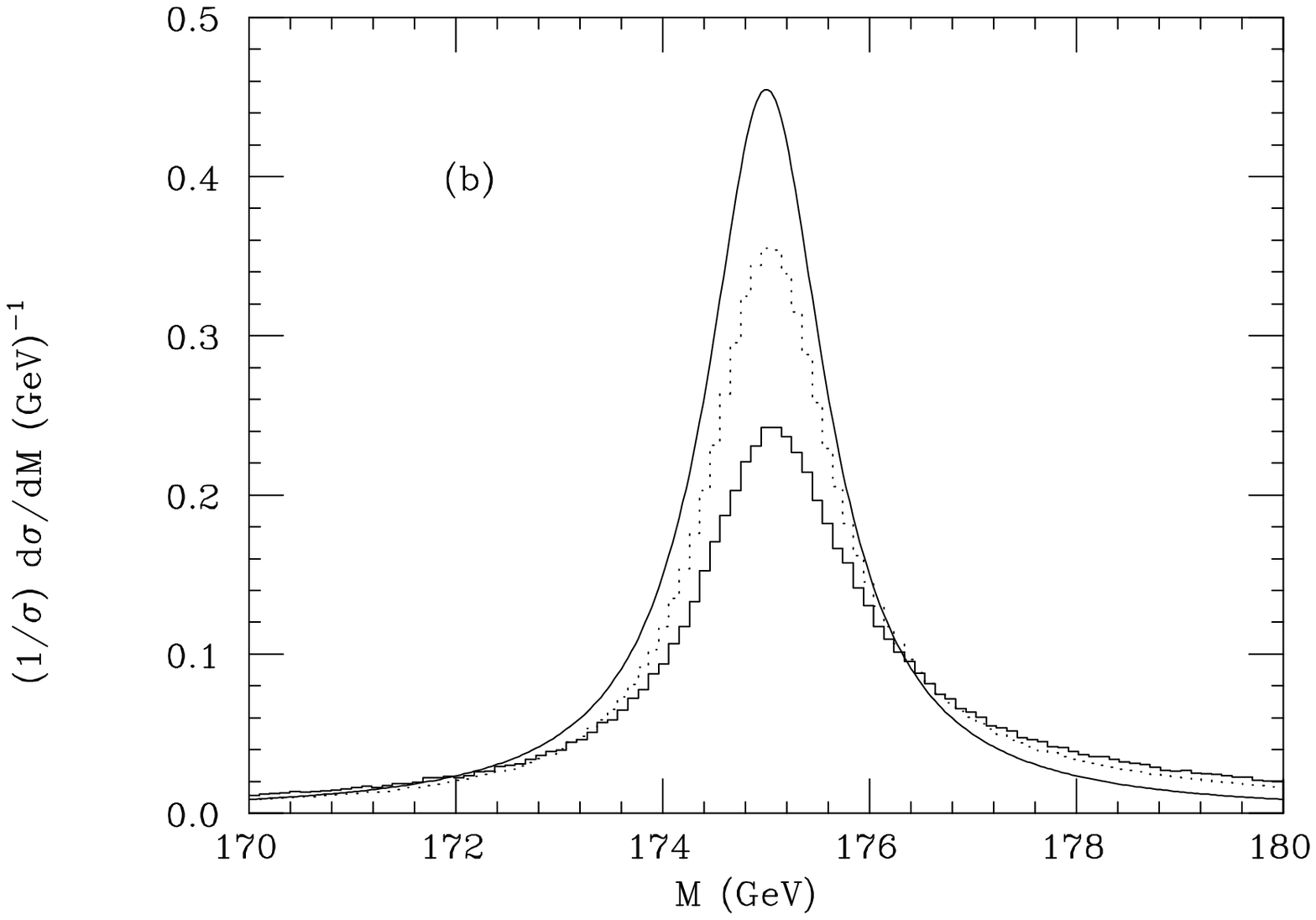}}
\vskip-4cm
\vskip6pt
\baselineskip=12pt
Fig.~9: Top mass reconstruction distributions without energy smearing
of the final-state partons for $\sqrt{s}=400$ GeV
(a) in the all-jet mode and
(b) in the lepton+jets mode.  In both plots the dotted histogram
is at tree-level, the solid histogram is at ${\cal O}(\alpha_s)$, and
the smooth curve is the original Breit-Wigner distribution.
\end{figure}

\newpage

\begin{figure}
\vskip-4cm
\epsfysize=16cm
\centerline{\epsffile{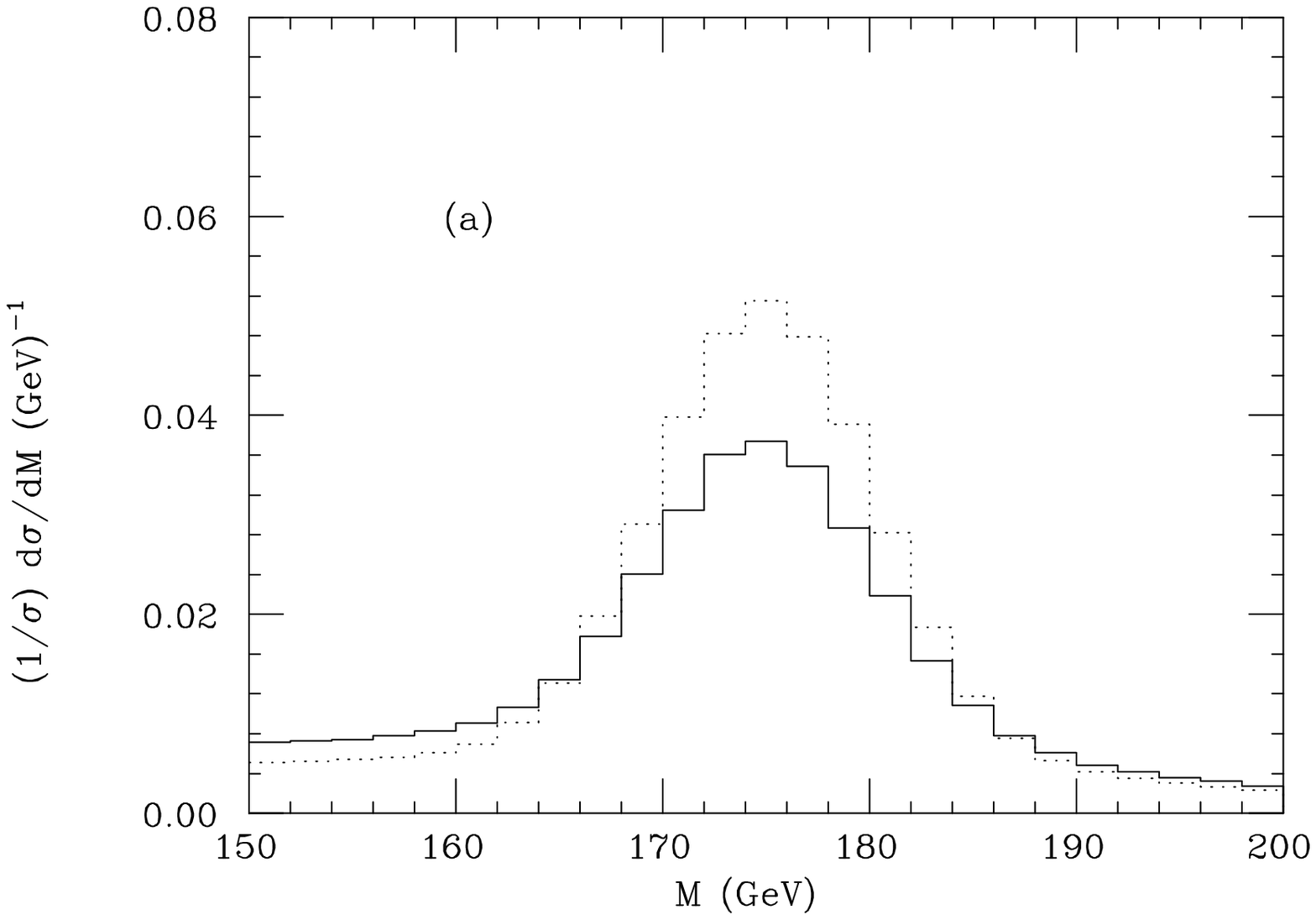}}
\vskip-4cm
\vskip-4cm
\epsfysize=16cm
\centerline{\epsffile{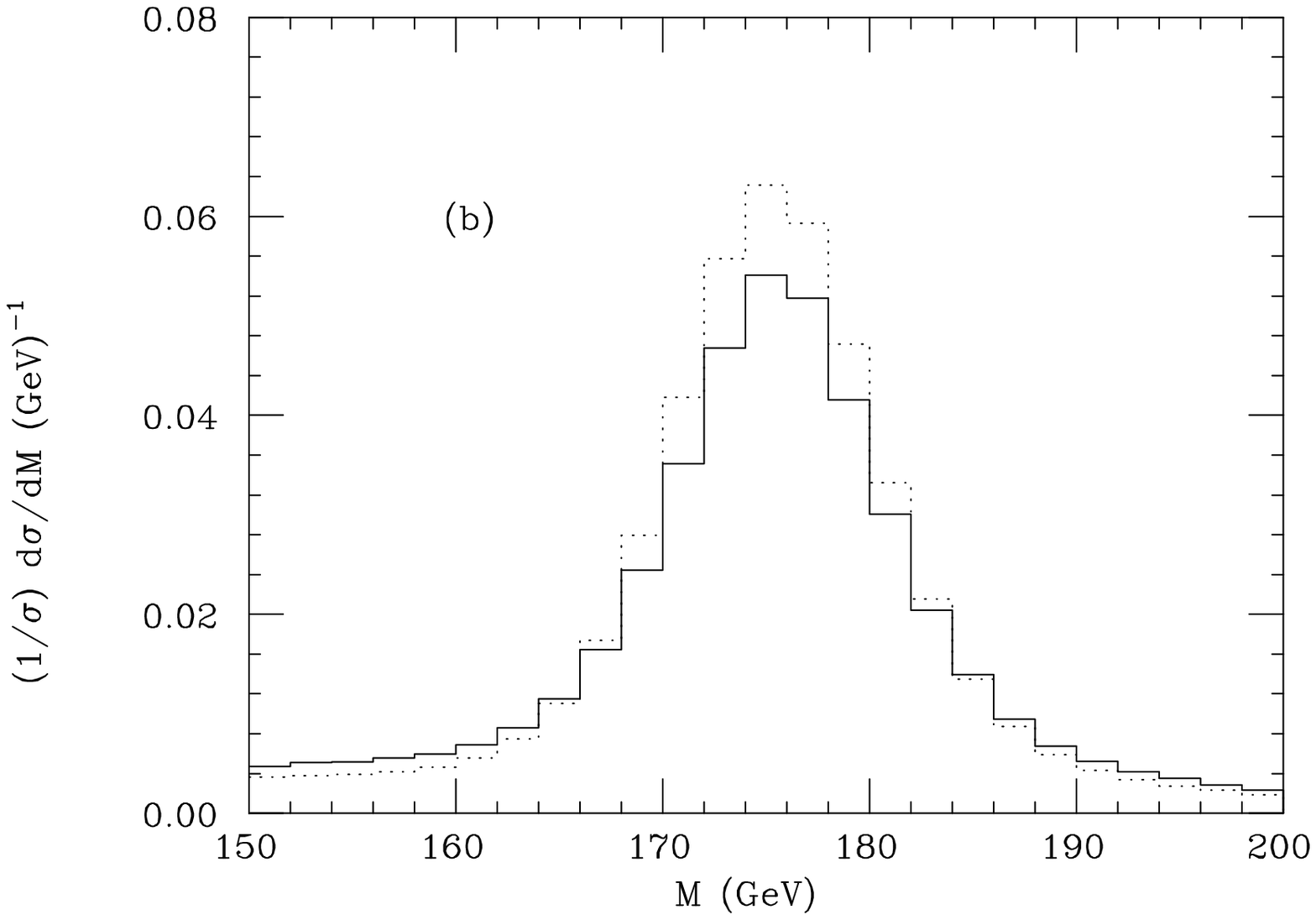}}
\vskip-4cm
\vskip6pt
\baselineskip=12pt
Fig.~10: Top mass reconstruction distributions with energy smearing
of the final-state partons for $\sqrt{s}=400$ GeV
(a) in the all-jet mode and
(b) in the lepton+jets mode.  In both plots the dotted histogram
is at tree-level and the solid histogram is at ${\cal O}(\alpha_s)$.
\end{figure}


\begin{thebibliography}{99}
\bibitem{cluster}
The jet-clustering used here is labeled algorithm E in
S. Bethke et al.,
{\it Nucl. Phys. } {\bf B370} (1992) 310.
At the order to which we are working the
differences between algorithms is not significant.
\end{thebibliography}
\end{document}